\documentclass[preprint,12pt]{elsarticle}
\usepackage[T1]{fontenc}
\usepackage[latin9]{inputenc}
\usepackage{amsmath}
\usepackage{amsthm}
\usepackage{graphicx}
\usepackage{float}
\usepackage{subcaption}
\usepackage{algorithm}
\usepackage{algpseudocode}
\usepackage{setspace}
\usepackage{cuted}

\theoremstyle{definition}
 \newtheorem{example}{\protect\examplename}
\theoremstyle{plain}
\newtheorem{lem}{\protect\lemmaname}
\theoremstyle{definition}
\newtheorem{defn}{\protect\definitionname}
\theoremstyle{plain}

\theoremstyle{plain}
\newtheorem{thm}{\protect\theoremname}
\DeclareMathOperator*{\argmax}{argmax}

\usepackage{babel}

\makeatother
\date{}
\usepackage{babel}
\providecommand{\definitionname}{Definition}
\providecommand{\examplename}{Example}
\providecommand{\lemmaname}{Lemma}
\providecommand{\propositionname}{Proposition}
\providecommand{\theoremname}{Theorem}

\usepackage{caption}

\allowdisplaybreaks
\begin{document}

\begin{frontmatter}
\title{Approximation Schemes for Age of Information Minimization in UAV Grid Patrols}

\author[label1]{Weiqi Wang}
\affiliation[label1]{Department of Physics and Astronomy, University of Victoria,
3800 Finnerty Rd, Victoria, BC V8P 5C2, Canada. Email: weiqiwang@uvic.ca}
\author[label2]{Jin Xu}
\affiliation[label2]{School of Management, Huazhong University of Science and Technology, Wuhan 430074, PR China. Corresponding author. Email: xu_jin@hust.edu.cn}

\begin{abstract}
Motivated by the critical need for unmanned aerial vehicles (UAVs) to patrol grid systems in hazardous and dynamically changing environments, this study addresses a routing problem aimed at minimizing the time-average Age of Information (AoI) for edges in general graphs. We establish a lower bound for all feasible patrol policies and demonstrate that this bound is tight when the graph contains an Eulerian cycle. For graphs without Eulerian cycles, it becomes challenging to identify the optimal patrol strategy due to the extensive range of feasible options. Our analysis shows that restricting the strategy to periodic sequences still results in an exponentially large number of possible strategies. To address this complexity, we introduce two polynomial-time approximation schemes, each involving a two-step process: constructing multigraphs first and then embedding Eulerian cycles within these multigraphs. We prove that both schemes achieve an approximation ratio of 2. Further, both analytical and numerical results suggest that evenly and sparsely distributing edge visits within a periodic route significantly reduces the average AoI compared to strategies that merely minimize the route travel distance. Building on this insight, we propose a heuristic method that not only maintains the approximation ratio of 2 but also ensures robust performance across varying random graphs.
\end{abstract}

\end{frontmatter}

\section{Introduction}

Unmanned aerial vehicles (UAVs) present a novel approach for grid surveillance and damage detection across various critical infrastructures, including power grids \cite{zhang2021optimal}, highways \cite{outay2020applications}, and pipelines \cite{Thomson2023Drone}. These grids, often located in remote and hard-to-access areas, greatly benefit from UAV technology, which facilitates the rapid acquisition of crucial data for operational management. For instance, UK Power Networks now utilize drones to swiftly identify disruptions and determine the causes of power outages \cite{Austin2024Drone}. Moreover, these grids are especially susceptible to more hazardous and dynamic environments during natural disasters (e.g., wildfires, earthquakes), extreme weather conditions (e.g., blizzards, low temperatures), and accidents. Unlike routine surveillance, UAVs must repeatedly patrol the grids in these situations to ensure timely detection of any damage caused by the dynamically changing environment.

The Age of Information (AoI) is a metric measuring system operators' freshness of knowledge about the dynamic environment (see \cite{kaul2012real,sun2019age,yates2021age}). Assuming that the sensed information can be transmitted and used for immediate decision-making, the AoI at a specific location is defined as the elapsed time since the location was last visited. A high AoI indicates a long absence of surveillance at that location, which is particularly risky in rapidly changing environments where damage may occur and remain unnoticed due to infrequent monitoring (see \cite{Ornee2021sampling,dong2019age,chen2021information}). Consequently, AoI is a critical metric that system operators aim to minimize to ensure timely monitoring, diagnosis, and preventative maintenance of each grid. Therefore, the patrolling policy for UAVs must be carefully designed to ensure frequent and timely visits to all parts of the grid.

In this work, we address the patrolling problem in general graphs, with the goal of minimizing the time-average AoI for graph edges. Minimizing the time-average AoI necessitates that the UAV repeatedly patrols the graph, presenting new research challenges in routing policy design. Traditional routing problems primarily focus on minimizing total travel distances or costs. For instance, in the well-known Traveling Salesman Problem (TSP), all the nodes of the graph must be visited with a minimum traveling distance \cite{flood1956traveling}. Similarly, in the Chinese Postman Problem (CPP), all the edges must be traversed with the total distance minimized \cite{eiselt1995arc}. In contrast, our scenario requires each grid location to be visited repeatedly and frequently to maintain up-to-date information. This requirement significantly differs from classical problems that require each node or edge to be visited at least once but do not require revisits.

Previous studies on AoI in UAV patrols primarily focused on minimizing the AoI over the nodes in graphs. For example, Zhu et al. \cite{2023Botao} and Ren et al. \cite{2023Mingyuan} examined the AoI minimization problem where UAVs collect data from sensor nodes within a graph. Luo et al. \cite{luo2023trajectory} and Liang et al. \cite{liang2023age} focused on optimizing UAV trajectories to minimize AoI over sensor nodes within a graph, considering additional parameters like localization accuracy and bandwidth to ensure information freshness. Hazarika and Rahmati \cite{2022Hazarika} investigated a multi-tier UAV sensing problem for discrete targets within a graph, considering AoI as one of the objectives. Luo et al. \cite{luo2022uav} proved that the AoI minimization problem for nodes over grid-graphs is NP-hard and proposed polynomial-time approximation schemes to solve the problem. Other node-specific studies considering AoI can be found in \cite{abd2019deep,tong2020deep,yi2020deep,abd2018average,hu2020aoi,liu2018age}. These studies mainly focused on the AoI of discrete nodes within a graph, without investigating the AoI of edges, which is crucial in the grid-patrolling scenario. 

The distinctions between classic metrics and AoI, as well as between node-AoI and edge-AoI, introduce significant challenges in the design and analysis of policies that minimizes AoI for edges. Existing research lacks frameworks for evaluating the time-average AoI for edges within a graph that is continuously and repeatedly patrolled by UAVs. Furthermore, the complexity of the AoI minimization for edges remains unknown. Given that the time-average AoI is a long-term metric, the range of potential patrol policies could be vast, including non-periodic ones. This complexity makes finding an optimal patrol strategy particularly challenging.

In this research, we address the outlined challenges by providing approximation schemes and theoretical analysis. The contributions of our work are summarized as follows.
\begin{itemize}
    \item \emph{(Modeling of AoI for edges):} We define the time-average AoI for the first time using the Lebesgue measure, providing a rigorous basis for further analysis. We also demonstrate the equivalence of this definition to a more intuitive Riemann sum form. Additionally, we provide an algorithm to effectively compute the AoI of a given route based on these definitions.
    
    \item \emph{(Problem complexity and theoretical bound):} We derive a lower bound of AoI applicable to all feasible patrol policies and prove that this bound is tight when Eulerian cycles exist within the graph. When Eulerian cycles do not exist, we show that the number of feasible policies within this set grows exponentially with the number of edges and nodes.

    \item \emph{(Approximation schemes and performance guarantee):} Furthermore, we develop two approximation schemes that transform the original graph into an Eulerian graph by adding auxiliary edges, and use this intermediate Eulerian structure to derive efficient patrol routes. These schemes are demonstrated to achieve an approximation ratio of 2. Through both analytical results and numerical studies, we reveal that optimally minimizing AoI involves distributing patrols sparsely and evenly across edges within a finite route rather than simply minimizing travel distance. This result highlights the fundamental difference between the AoI-minimizing routing and the classic routing problems. Building on this insight, we have also crafted a heuristic algorithm that not only offers an approximation ratio 2 but also exhibits decent average performance.
\end{itemize}

The rest of the paper is organized as follows: We introduce the AoI
minimization problem and discuss its properties in Section \ref{sec:The-AoI-minimization-problem}.
Section \ref{sec:approximation schemes} introduces the approximation
schemes. We provide a heuristic approach and perform numerical studies
in Section \ref{sec:heutistic-and-numerical}. Finally, we summarize our findings and discuss potential avenues for future research in Section \ref{sec:conclusions-and-future}.

\section{The AoI-Minimization Problem \label{sec:The-AoI-minimization-problem}}

\begin{figure}
\centering
\includegraphics[scale=0.7]{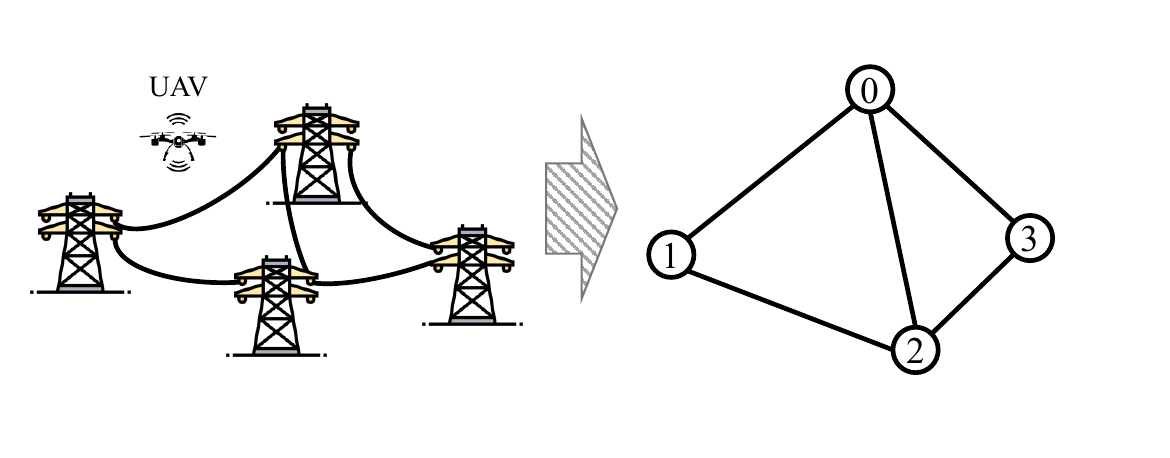}
\caption{Demonstrative graph of the grid patrol}\label{fig:UAV}
\end{figure}

In this section, we first define the AoI of edges on a general connected
graph and then provide several preliminary results. We assume that the network that a UAV needs to patrol can be modeled as a connected simple graph $\mathcal{G}=(V,E)$ where any two nodes can be connected by only one edge, as shown in Figure \ref{fig:UAV}. Without
loss of generality, we index the nodes in the graph by $\mathcal{N}=(0,1,...,N-1)$,
and each age connecting nodes $i$ and $j$ as $\{E_{i,j}\}$. We assume
that the UAV traverses along edges at a unit speed. 
All the edges must be patrolled periodically so that the UAV can spot any errors as soon as possible. As the UAV can patrol the edges in various patterns,
we restrict our discussion to a special type of policies $\mathcal{F}.$
For any patrol policy $\psi$ within $\mathcal{F}$, all edges are traversed
by the UAV without changing directions halfway, and the UAV never
halts while traversing. As our main focus in this paper is to investigate how to design patrol strategies to minimize AoI, we maintain a rather ideal and mathematical model by excluding the charging and maintenance activities of the UAV, considering only the case of a single UAV patrolling the network.

The objective of UAV patrolling is to minimize the time-average Age of Information over all the edges within the graph $\mathcal{G}$, which we will formally define in the following. The temporary age of location $x$ (a point on the edge) at time $t$ is defined as $\Delta(x,t)=t-U(x,t)$, where $U(x,t)$ is the most recent time that this point
was visited. Note that once the UAV arrives at location $x$, the temporary
age at location $x$ drops to zero. This captures the scenario where the UAV senses the information for the grid and transmits it to the operator immediately, with negligible data transmission time. The age thus measures the time
span since the last time the UAV visited the place. A large age indicates
that the UAV has not visited this location for a long period since the last visit, making it likely that a grid defect has occurred at this location but has not yet been detected by the UAV.

Based on the temporary AoI for each location, we now define the time-average AoI for the graph $\mathcal{G}=(V,E)$. We let $\mathcal{B}$ be the Borel $\sigma$-algebra defined on the set $X=\big\{ x|x\in\cup E_{i,j}\big\}$ containing all points on the graph, and we
define the measure on any $Y\in \mathcal{B}$ as 
\begin{equation}
l(Y)=\inf\left\{\sum_{k=1}^{\infty}\mu(S_{k})\biggr|Y\subset\cup_{k=1}^{\infty}S_{k}\right\},\label{eq:2-1}
\end{equation}
where $\mu(S_{k})$ represents the length of the open interval $S_{k}\subset X$. Here, the length of the open interval $S_k$, denoted as $\mu(S_k)$, is defined as the sum of the distances between the endpoints of $S_k$ on all edges.
Let $Y_{u}(t)=\{x\in\mathcal{B},\Delta(x,t)>u\}$ be the
set from $\mathcal{B}$ that contains all the points with temporary
age greater than $u$. The total age of all the edges at a particular time $t$ is represented as $\int_{u=0}^{\infty}l(Y_{u}(t))du$. We can then define the time-average AoI for graph $\mathcal{G}$ under patrolling policy $\psi$ as 
\begin{eqnarray}
\overline{\Delta}(\mathcal{G};\psi) & = & \lim_{T\rightarrow\infty}\frac{1}{T}\int_{t=0}^{T}\int_{u=0}^{\infty}l(Y_{u}(t))dudt.\label{eq:2-2}
\end{eqnarray}
To facilitate our discussion, we assume throughout our paper that the temporary age at time $t$ is finite almost everywhere, i.e., $l(Y_{\infty}(t))=0$. We then have 
$$\int_{u=0}^{\infty}l(Y_{u}(t))du   = \int_{x\in\cup E_{i,j}}\Delta(x,t)dx.$$
This conversion effectively transforms a Lebesgue integral into a more intuitive Riemann sum form by directly integrating the age over all points on the graph. Integrating $l(Y_{u}(t))$ with respect to $u$ from 0 to $\infty$ sums the lengths of the intervals where the age exceeds each possible value $u$. This process counts the total age over the edges, analogous to summing areas under a curve in the Riemann integral. Thus, we can
rewrite the time-average AoI in Equation \eqref{eq:2-2} as 
\begin{equation}
\overline{\Delta}(\mathcal{G}; \psi) =  \lim_{T\rightarrow\infty}\frac{1}{T}\int_{t=0}^{T}\int_{x\in\cup E_{i,j}}\Delta(x,t)dxdt.\label{eq:2-3}
\end{equation}
Both Equations \eqref{eq:2-2} and \eqref{eq:2-3} will be useful for our analysis later. We provide a demonstrative example in the following on how to compute the time-average AoI on a single edge using Equation  \eqref{eq:2-3}. 
\begin{example}
\label{ex:Consider-the-scenario} In this example, we consider a basic scenario where the graph $\mathcal{G}$ only has two nodes $\{0,1\}$ and a single edge $E_{0,1}$ with length $a$.
The UAV patrols the line back and forth at a unit speed. Consequently, the
UAV visit node $0$ at times $0,2a,4a,...$, and visit node 1
at times $a,3a,5a,...,$, and so on. A snapshot
of the temporary age of the edge $E_{0,1}$ is given in Figure \ref{fig:A-snapshot-of-1}, where the interval $[0,a]$ on the x-axis represents edge $E_{0,1}$. We assume $\Delta(x,0)<\infty$ for all $x\in [0,1]$. Since the accumulated AoI within the intervals $[k\cdot a,(k+1)\cdot a]$ with $k\in \mathbf{Z}_{+}$ are identical, we consider a time $\tau+2a\in[2a,3a]$ where the UAV
is at the location $\tau$ and traversing towards node 1. 
The age integrated over $E_{0,1}$ at time $\tau+2a$ is given by: 
\begin{align*}
\int_{x=0}^{a}\Delta(x,\tau+2a)dx & = \frac{\tau^{2}}{2}+\frac{2\tau+a+\tau}{2}(a-\tau) \\
& =  -\tau^{2}+\tau a+\frac{a^{2}}{2}.
\end{align*}
We then consider the accumulated AoI for edge $E_{0,1}$ during the time interval $[2a,3a]$:
$$
\int_{t=2a}^{3a}\int_{x=0}^{a}\Delta(x,t)dxdt = \!\int_{\tau=0}^{a}(-\tau^{2}+\tau a+\frac{a^{2}}{2})d\tau
  \!= \! \frac{2}{3}a^{3}.
$$
As the accumulated AoI at each round of visit is identical, based on Equation \eqref{eq:2-3}, we have
\begin{equation*}
\overline{\Delta}(\mathcal{G};\psi)  =\frac{1}{a}\int_{t=2a}^{3a}\int_{x=0}^{a}\Delta(x,t)dxdt =  \frac{2}{3}a^{2}.
\end{equation*}

\begin{figure}[!t]
\centering
\includegraphics[scale=0.4]{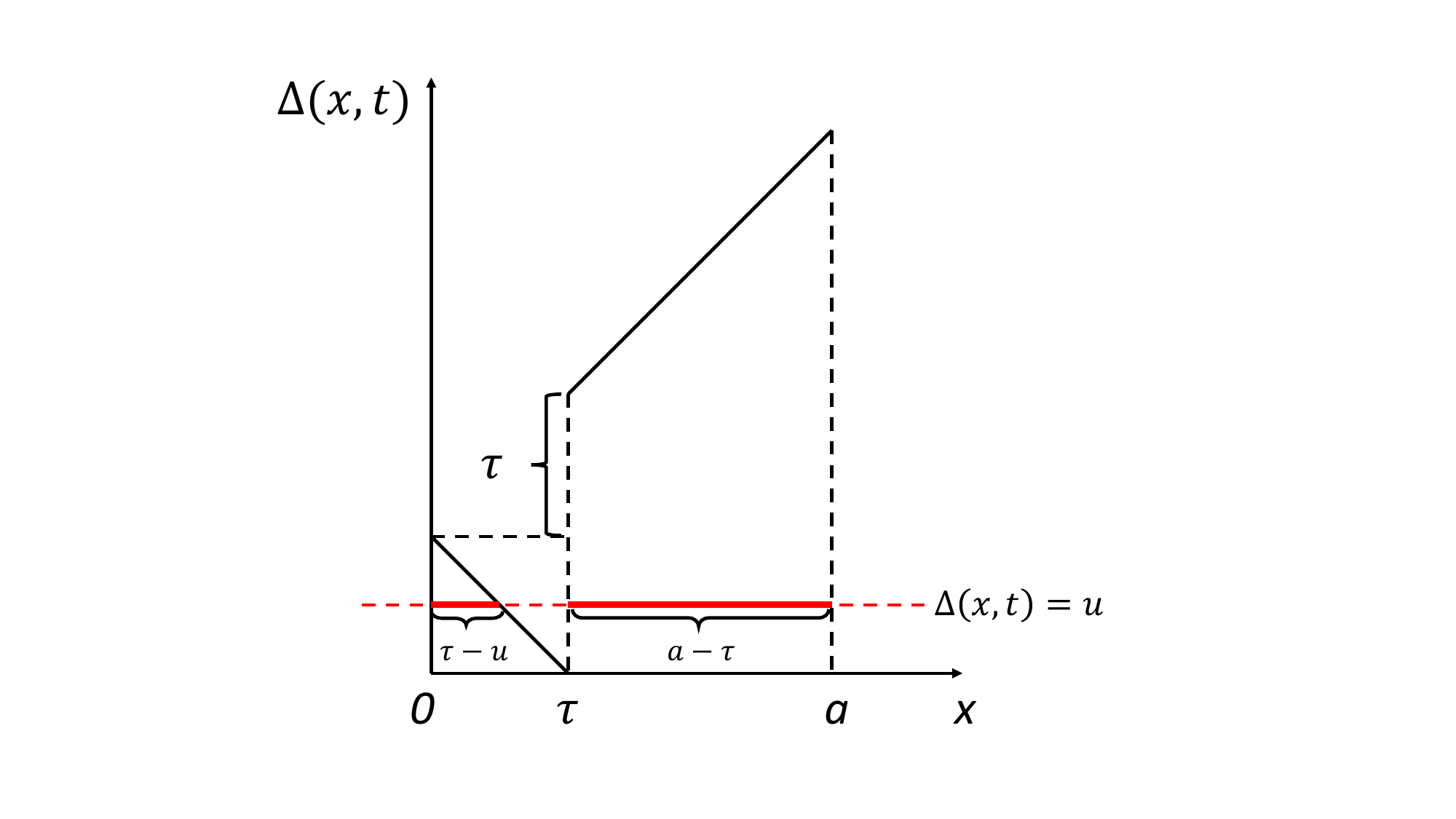}\caption{A snapshot of the AoI on interval $[0,a]$ at $t=\tau+2a$ ($0<\tau<a$). }\label{fig:A-snapshot-of-1}
\end{figure}
\end{example}

The example above illustrates how to compute the time-average AoI on a single edge. Computing the AoI for multiple edges is more complex. We will provide an algorithm for AoI computation in Section \ref{sec:approximation schemes}. 

Our aim is to find a policy from $\mathcal{F}$ that minimizes $\overline{\Delta}(\mathcal{G})$. However, the number of policies in $\mathcal{F}$ is infinite and these policies can even be non-periodic. We will show later that even if we restrict our policy to a periodic policy set, the number of feasible policies can still be exponential. This makes searching for the optimal policy challenging. Nevertheless, we can derive the lower bound of the AoI for all the feasible policies, including non-periodic ones, in the following lemma. 

\begin{lem}
\label{lem:For-any-connected}For any connected graph $\mathcal{G}=(V,E)$,
we always have $\overline{\Delta}(\mathcal{G})\geq\frac{1}{2}l(E)^{2}$ with $l(E)=l(\cup E_{i,j})$.
The lower bound can be reached when there exists an Eulerian cycle
within $\mathcal{G}$. 
\end{lem}
\proof
Suppose that at finite time $t>0$, all the edges within $E$ have
been traversed at least once since time 0. We consider the
snapshot of the age at time $t$. Since $l(Y_{u}(t))$ is the total
edge length that has AoI greater than $u$ at time $t$, we have 
\begin{equation}
l(Y_{u}) \geq l(E)-u,\label{eq:2-4}
\end{equation}
where $l(E)$ is the length of all the edges within $E$. The reason for Equation \eqref{eq:2-4} is that the UAV traverses
the edges with a constant speed, so the length of the edges with temporary age smaller than $u$ is at most $u$. 
We then have by Equation \eqref{eq:2-2} that 
\begin{align*}
\overline{\Delta}(\mathcal{G}) = & \lim_{T\rightarrow\infty}\frac{1}{T}\int_{t=0}^{T}\int_{u=0}^{\infty}l(Y_{u}(t))dudt\\
  \geq & \lim_{T\rightarrow\infty}\frac{1}{T}\int_{t=0}^{T}\int_{u=0}^{l(E)}(l(E)-u)dudt=\frac{1}{2}l(E)^{2}.
\end{align*}

When there is an Eulerian cycle within $\mathcal{G}$, the UAV can
traverse the graph along the Eulerian cycle periodically, and each edge is traversed exactly once within each period. We thus have $l(Y_{u}(t))=l(E)-u$ and
$\overline{\Delta}(\mathcal{G})=\frac{1}{2}l(E)^{2}$, and the lower bound is tight in this case.
\endproof

Note that whether a graph has an Eulerian cycle can be easily checked
by examining the degrees of the nodes. If the degree of each node within $\mathcal{G}$ is even, then an Eulerian cycle always exists (see Theorem 6.2 of \cite{wilson2010introduction}). However,
if no Eulerian cycle exists, the AoI minimization problem becomes more challenging to solve. In the following section, we will provide analysis and approximation schemes for this scenario.

\section{Approximation Schemes\label{sec:approximation schemes}}

In this section, we investigate the scenario where an Eulerian
cycle does not exist in the graph. We restrict our discussion on a policy set with periodic policies, and introduce the way of computing AoI for a specific route in Section \ref{subsec:route}. We then provide the lower and upper bounds of the AoI for these periodic policies in Sections \ref{subsec:Lower-Bound-of} and \ref{subsec:upper-bound}. Based on these bounds, we propose two approximation schemes in Sections \ref{subsec:dup} and \ref{sec:CPP} and prove their approximation ratios.

\subsection{Periodic Policy Set $\mathcal{F}_{1}$ and Routes}\label{subsec:route}

We introduce a subset $\mathcal{F}_{1}\subset\mathcal{F}$ that
contains all the policies that periodically traverse edges, and all
the edges are traversed at least once and at most twice within each period. Since
periodic policies can be fully characterized by the sequence of edges
that the UAV traverses within a period, we define the concept of a \emph{route
}as follows. 
\begin{defn}\label{def;F_1}
A route $R \in \mathcal{F}_{1}$ is defined as a sequence of edges characterized
by the order and the directions of the visits. A route in $\mathcal{F}_{1}$ should traverse
all the edges at least once and at most twice before returning to
the starting node. 
\end{defn}

Since we assume our original grid is a simple graph $\mathcal{G}$, each edge can be uniquely characterized by the nodes it links.  We can thus characterize a route $R$ by a vector $R=(R[0],...,R[M])$, with each element $R[i]$ in the vector denoting the $i^{th}$ node that
it visits. A route will always start and end at the same node, so we have $R[0]=R[M]$. For instance, we can denote the route in Example \ref{ex:Consider-the-scenario} that
traverses node 0 and 1 back and forth as $(0,1,0)$,
where the first and last elements in the vector are the starting and
ending nodes. With a bit abuse of notation, we use $l(R)$ to denote the total distance traversed along $R$ within each period, i.e., $l(R)=\sum_{i=0}^{M-1}l(E_{R[i],R[i+1]})$. We denote the AoI of graph $\mathcal{G}$ under route $R\in\mathcal{F}_{1}$
as $\overline{\Delta}(\mathcal{G};R)$.

\begin{figure}
\centering
\includegraphics[scale=0.5]{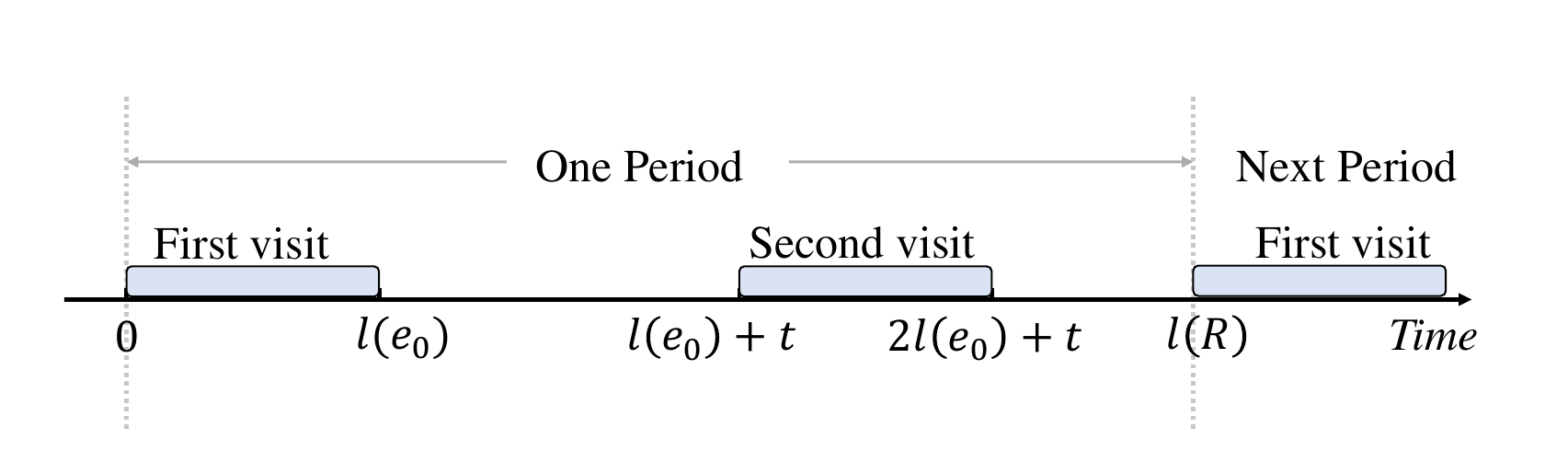}\caption{A demonstrative graph for the two visits of $e_0$ if $e_0$ is traversed twice within $R$}\label{fig:A-demonstrative-graph}
\end{figure}

We next introduce the method to calculate $\overline{\Delta}(\mathcal{G};R)$ for a given route $R$ by focusing on an arbitrary
edge $e_{0}\in E$. We first discuss the scenario where the edge $e_0$ is traversed twice within $R$. We assume that the UAV starts to patrol
edge $e_{0}$ for the first time from location $x=0$ at time 0, and finishes patrolling at location $x=l(e_0)$ at time $l(e_{0})$.
The UAV starts to patrol edge $e_{0}$ again at time $l(e_{0})+t$,
and finishes patrolling at time $2l(e_{0})+t$. A demonstrative graph of the two visits of edge $e_0$ is given in Figure \ref{fig:A-demonstrative-graph}. 

We then compute the accumulated AoI for edge $e_{0}$ during the epoch $[l(e_{0}),2l(e_{0})+t]$. From time $l(e_{0})$ to time $l(e_{0})+t$, edge $e_0$ is not visited, and its accumulated AoI during this period is given as 
\begin{align}\label{eq:3-1}
& \int_{\tau=l(e_{0})}^{l(e_{0})+t}\int_{x\in e_{0}}\Delta(x,\tau) dxd\tau \nonumber \\
 = & \int_{\tau=l(e_{0})}^{l(e_{0})+t} \int^{l(e_0)}_{x=0}\left(\tau - l(e_0) +(l(e_0)-x)\right) dxd\tau \nonumber \\
  = & \frac{1}{2} t \cdot l(e_0) \left( l(e_0)+ t\right).
\end{align}

\begin{figure}[!t]
\centering \begin{subfigure}{.4\textwidth} \includegraphics[width=1\linewidth]{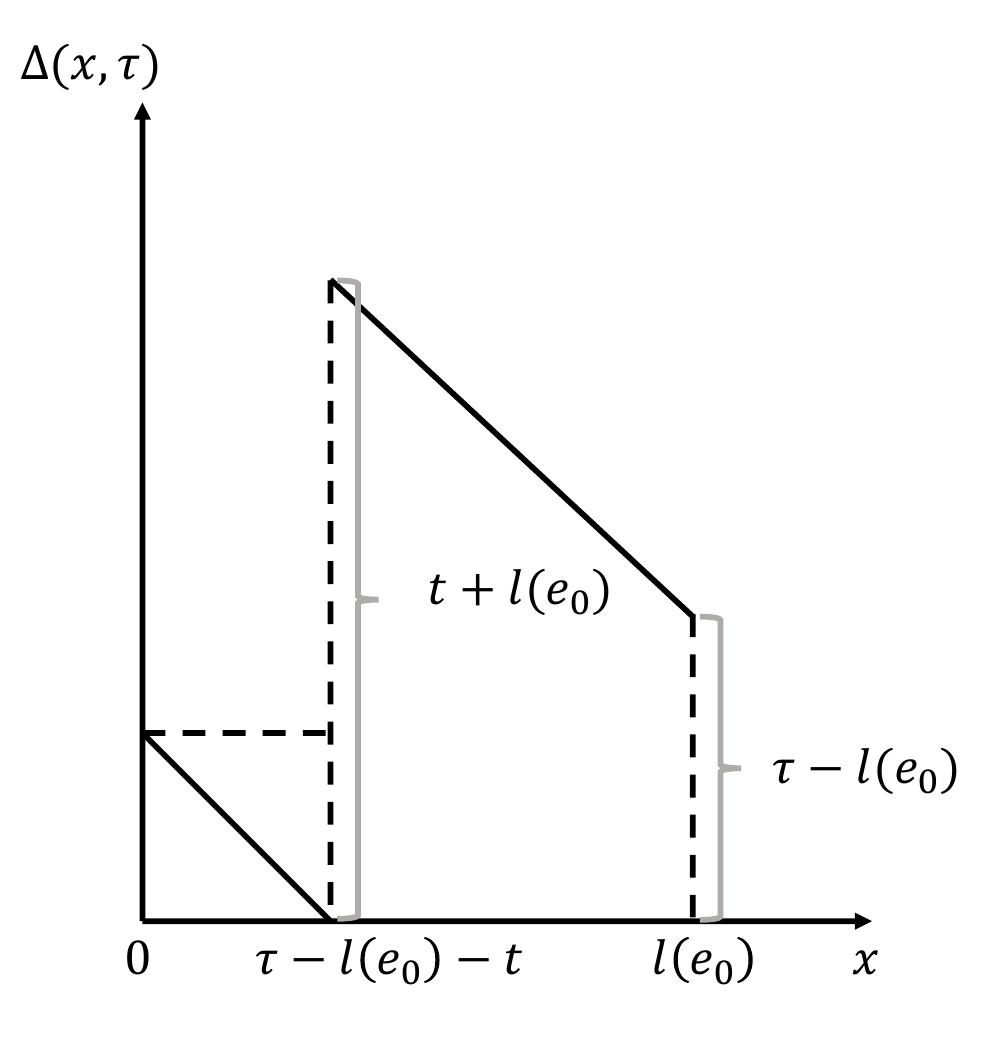}
\subcaption{Two visits from the same direction} \end{subfigure} \begin{subfigure}{.4\textwidth}
\includegraphics[width=1\linewidth]{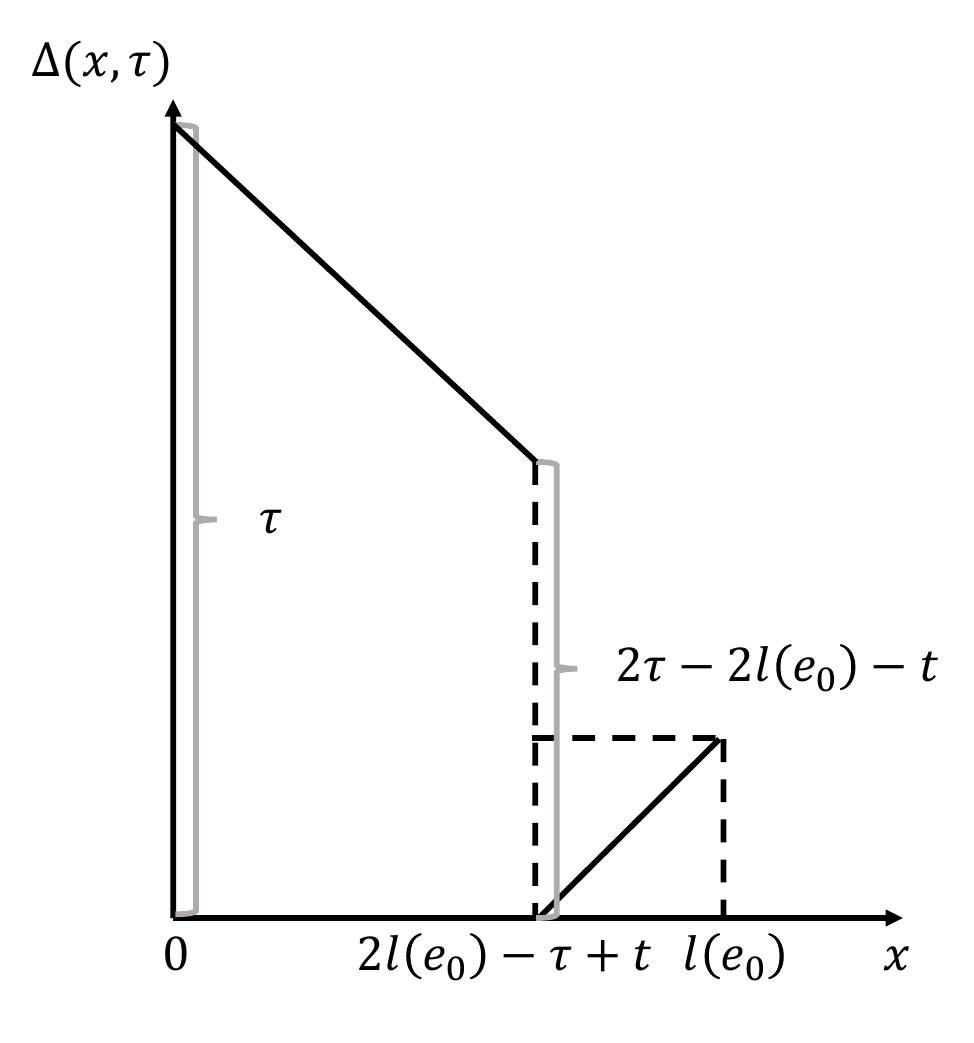}
\subcaption{Two visits from opposite directions} \end{subfigure} \caption{A snapshot of the age of edge $e_0$ at time $\tau$}
\label{Fig:snapshot-at-tau} 
\end{figure}

From time $l(e_{0})+t$ to time
$2l(e_{0})+t$, the UAV patrols edge $e_0$ for the second time. If the UAV traverses $e_{0}$ from the same direction
at times 0 and $l(e_{0})+t$ (i.e., from $x=0$ to $x=l(e_0)$), then the location of the UAV at time $\tau\in[l(e_{0})+t,2l(e_{0})+t]$  is $\tau - l(e_0) - t$. A snapshot of the temporary age at time $\tau$ is given in Figure \ref{Fig:snapshot-at-tau}, where we consider edge $e_0$ as the interval $[0,l(e_0)]$ in the x-axis. Then, we calculate the accumulated AoI during this period as
\begin{align}\label{eq:3-2}
 &   \int_{\tau=l(e_{0})+t}^{2l(e_{0})+t}\int_{x\in e_{0}}\Delta(x,\tau)dxd\tau \nonumber\\
  = & \int_{\tau=l(e_{0})+t}^{2l(e_{0})+t}\Big(\int_{x=0}^{\tau-l(e_{0})-t}+\int_{x=\tau-l(e_{0})-t}^{l(e_{0})}\Big)\Delta(x,\tau)dxd\tau \nonumber \\ 
  = & \int_{\tau=l(e_{0})+t}^{2l(e_{0})+t}\frac{1}{2}\Big(\tau-l(e_{0})-t\Big)^{2}d\tau  + \int_{\tau=l(e_{0})+t}^{2l(e_{0})+t}\frac{1}{2}\Big(2l(e_{0})+t-\tau\Big)\Big(t+\tau\Big)\nonumber d\tau \\
  = & \frac{1}{2} l(e_0)^2 \left(l(e_0) +t\right).
\end{align}

If the direction that the UAV traverses $e_{0}$ for the second time is opposite to that of the first time, i.e.,  the second visit is from $x=l(e_0)$ to $x=0$, the UAV is at location $2l(e_0)-\tau +t$ at time $\tau$. A snapshot of the temporary age at time $\tau$ is given in Figure \ref{Fig:snapshot-at-tau}(b). We can then compute the accumulated AoI as: 
\begin{align}\label{eq:3-3}
   & \int_{\tau=l(e_{0})+t}^{2l(e_{0})+t}\int_{x\in e_{0}}\Delta(x,\tau)dxd\tau \nonumber\\
  = & \int_{\tau=l(e_{0})+t}^{2l(e_{0})+t}\frac{1}{2}\Big(\tau-l(e_{0})-t\Big)^{2}d\tau \nonumber \\
  & +\int_{\tau=l(e_{0})+t}^{2l(e_{0})+t}\frac{1}{2}\Big(2l(e_{0})+t-\tau\Big)\Big(3\tau-t-2l(e_0)\Big)\nonumber d\tau \\
  = & \frac{1}{2} l(e_0)^2 \Big(\frac{4}{3}l(e_0) +t\Big).
\end{align}

We can thus compute the accumulated AoI during the epoch $[l(e_{0}),2l(e_{0})+t]$ as $\int_{\tau=l(e_{0})}^{2l(e_{0})+t}\int_{x\in e_{0}}\Delta(x,\tau)dxd\tau$
based on Equations \eqref{eq:3-1} and \eqref{eq:3-2} or Equations \eqref{eq:3-2} and \eqref{eq:3-3}. Note that the
accumulated AoI during the time interval $[l(e_{0}),2l(e_{0})+t]$ only depends
on the interval $t$ between two visits and the edge length $l(e_{0})$. Similarly, we can compute the accumulated AoI during the epoch $[2l(e_{0})+t,l(R)]$, and then obtain the accumulated AoI for $e_0$ over the entire period as $\int_{\tau=0}^{l(R)}\int_{x\in e_{0}}\Delta(x,\tau)dxd\tau$. If $e_{0}$ is traversed only once within $R$, we can compute the accumulated AoI using Equations \eqref{eq:3-1} and \eqref{eq:3-2} with $t=l(R)-l(e_0)$, as the edge $e_0$ is always traversed from the same direction.  Based on this idea, we can compute the accumulated AoI for all the edges in $E$, and the algorithm is summarized in Algorithm \ref{alg:Compute-the-AoI}.

We explain the steps in Algorithm \ref{alg:Compute-the-AoI} as follows. Line 2 of Algorithm \ref{alg:Compute-the-AoI} creates an augmented route $R^{a}$ by repeating the original route $R$. Note that route $R$ is traversed exactly twice within $R^{a}$. The purpose of creating this augmented route is to obtain the time intervals between visits to each edge. Lines 10 and 11 show that if the edge is traversed in the same direction in two adjacent visits, then we compute the accumulated AoI based on Equations \eqref{eq:3-1} and \eqref{eq:3-2}. If it is traversed in opposite directions, we compute the accumulated AoI based on Equations \eqref{eq:3-1} and \eqref{eq:3-3}, as shown in Lines 13 and 14. Line 19  shows that once the accumulated AoI of all the edges is summed, we average it on the length of the route $R$ (the total distance traveled within one period) to obtain the time-average AoI for the entire graph. 

\begin{algorithm}
\footnotesize{
\begin{algorithmic}[1] \State{\textbf{Input:} Graph $\mathcal{G}=(V,E)$
and route $R\gets(R[0],...,R[M])$. }

\State{Create an augmented route $R^{a} \gets (R[0],..., R[M-1],R[0]$ $,...,R[M])$, Accumulated AoI $\Delta\gets0$.
}

\For{$i\gets M$ to $2M-1$}

\State{Current edge $e\gets (R^a[i],R^a[i+1])$.}
\State{$t\gets 0$}

\For{$j\gets i$ down to $1$}

\State{$e_{1}\gets(R^a[j-1],R^a[j])$ and $e_{2}\gets(R^a[j],R^a[j-1])$. }

\If{$e_{1}\neq e$ and $e_{2}\neq e$}

\State{$t\gets t+l(e_{1})$}

\ElsIf{$e_{1}=e$}

\State{$\Delta\gets \Delta+\frac{1}{2}t^{2}\cdot l(e)+t\cdot l(e)^{2}+\frac{1}{2}l(e)^{3}$}

\State{Break}

\ElsIf{$e_{2}=e$}

\State{$\Delta \gets \Delta+\frac{1}{2}t^{2}\cdot l(e)+t\cdot l(e)^{2}+\frac{2}{3}l(e)^{3}$}

\State{Break}

\EndIf

\EndFor

\EndFor

\State{\textbf{Output:} The average AoI as $\overline{\Delta}(\mathcal{G};R)\gets\frac{\Delta}{l(R)}$} \end{algorithmic}
\caption{Compute the AoI under a fixed route \label{alg:Compute-the-AoI}}
}
\end{algorithm}

Having introduced the method of computing AoI for a fixed route $R$, we now characterize the number of routes in $\mathcal{F}_{1}$. Compared with the original policy set $\mathcal{F}$, the policy set $\mathcal{F}_1$ is more tractable. However, as shown in the following lemma, the number of routes
within $\mathcal{F}_{1}$ is still exponential. 

\begin{lem}
\label{lem:The-number-of} 
For a graph $\mathcal{G}=(V,E)$, the number of valid routes in $\mathcal{F}_{1}$
is lower bounded by $2^{1-n(V)+n(E)}$, where $n(V)$ is the number
of nodes and $n(E)$ is the number of edges. 
\end{lem}
\proof
For an arbitrary graph $\mathcal{G}$, regardless of whether an Eulerian
cycle exists in $\mathcal{G}$, one can always duplicate all the edges
in $\mathcal{G}$ and create an undirected multigraph $\mathcal{G}^{'}$. As the degrees
of all the nodes in $\mathcal{G}^{'}$ are even, 
Eulerian cycles always exist in $\mathcal{G}^{'}$. We then create a
policy set $\mathcal{F}_{2}$, where under any route $R\in \mathcal{F}_2$, the UAV traverses along an Eulerian
cycle in $\mathcal{G}^{'}$. Note that under any $R\in\mathcal{F}_{2}$,
each original edge in $\mathcal{G}$ is traversed exactly twice.
Therefore, we have $\mathcal{F}_{2}\subset\mathcal{F}_{1}$.

Since a route in $\mathcal{F}_{2}$ traverses all the edges in
$\mathcal{G}^{'}$, it will also visit all the nodes in each period.
Without loss of generality, we assume all the routes in $\mathcal{F}_{2}$ start from
and end at node 0. Routes in $\mathcal{F}_{2}$ differ in their sequences in visiting edges, as each route visits all edges twice but in potentially different orders and directions. As proven by \cite{punzi2022bounding}, the number
of edge-distinct Eulerian cycles that start and end at an arbitrary
node is lower bounded by $2^{1-n(V)+n(E)}$. This serves as a lower
bound for the number of routes in $\mathcal{F}_{2}$. Since  $\mathcal{F}_2 \subset \mathcal{F}_{1}$, the number of routes in $\mathcal{F}_1$ is also lower bounded by $2^{1-n(V)+n(E)}$. 
\endproof
As shown in Lemma \ref{lem:The-number-of}, when the number of edges
is significantly greater than the number of nodes, the policy $\mathcal{F}_{1}$
set can have an exponential number of policies. Consequently, searching for the
optimal policy within $\mathcal{F}_{1}$ could be intractable when
the number of edges is large. We aim to address this issue by providing approximation schemes. We next derive
the lower and upper bounds of the AoI for routes in $\mathcal{F}_{1}$ in the following subsections. These bounds will serve as the foundation for proving the approximation ratios of the schemes we propose later.

\subsection{Lower Bound of AoI in $\mathcal{F}_{1}$\label{subsec:Lower-Bound-of}}

In Lemma \ref{lem:For-any-connected}, we have introduced a lower bound
of AoI for all the policies in $\mathcal{F}$. Since $\mathcal{F}_{1}$
is a subset of $\mathcal{F}$, we aim to derive a tighter lower bound of
AoI under $\mathcal{F}_{1}.$ To achieve this, we introduce the concept of \emph{virtual
route} in the following.
\begin{defn}
A virtual route is defined as a sequence of edges characterized by the times and directions of the visits. Unlike a feasible route, a virtual route does not need to begin and end at the same node or be connected, and it can have overlapping visits (multiple visits to different edges at the same time stamp) as well as gaps (periods during which no edge is visited). 
\end{defn}
The AoI of virtual routes can be calculated in the same way as feasible routes, making them a useful tool for our analysis. In the following lemma, we derive the lower bound of the AoI based on the concept of virtual route. 
\begin{lem}
\label{lem:If-an-edge}For a graph $\mathcal{G} = (V, E)$, if an edge $e_{0}\in E$ is traversed twice
within a route $R$, then the AoI of edge $e_0$ in route $R$ is lower bounded by
that in the virtual route $R^{'}$, i.e., $\Delta(e_0;R^{'})\leq \Delta(e_0;R)$. In $R^{'}$, edge $e_{0}$
is traversed twice from the same direction, while other edges are traversed following the same sequence and direction as in $R$. 
\end{lem}
\proof
We consider the non-trivial case where $e_{0}$ is visited twice under $R$
and the directions of the two visits are different. Since the traversal time and direction of other edges are fixed, we only need to analyze the AoI for $e_{0}$. We assume the time intervals between the two visits of the edge
$e_{0}$ are $d_{1}$ and $d_{2}$, respectively. That is, once the UAV has traversed edge $e_0$, it takes time $d_1$ before traversing $e_0$ again. After this round of traverse, it takes the UAV another $d_2$ time before traversing $e_0$ again. Without loss of generality, we assume we start to travel $e_0$ at $t=0$. Applying Equations \eqref{eq:3-1} and \eqref{eq:3-2}, we compute the average AoI for $e_{0}$ under $R$ as
\begin{align}\label{eq:lem3-1}
\Delta(e_0;R) =&  \bigg(\int_{t=0}^{l(e_0)}+\int_{t=l(e_0)}^{l(e_0)+d_1}+\int_{t=l(e_0)+d_1}^{2l(e_0)+d_1} \nonumber \\
  + &\int_{t=2l(e_0)+d_1}^{2l(e_0)+d_1+d_2}\bigg)\int_{x\in e_{0}}\Delta(x,t;R)dxdt \nonumber\\
  = & \frac{4}{3}l(e_0)^3 + l(e_0)^2 (d_1+d_2) + \frac{1}{2}l(e_0)(d_1^2 +d_2^2 ). 
\end{align}

Similar to $R$, under $R^{'}$, the average AoI for $e_{0}$ during each period is
\begin{align}\label{eq:lem3-2}
&\Delta(e_0;R^{'}) \nonumber\\
 = & \bigg(\int_{t=0}^{l(e_0)}+\int_{t=l(e_0)}^{l(e_0)+d_1}+ \int_{t=l(e_0)+d_1}^{2l(e_0)+d_1} + \int_{t=2l(e_0)+d_1}^{2l(e_0)+d_1+d_2}\bigg)\int_{x\in e_{0}}\Delta(x,t;R^{'})dxdt\nonumber\\
= & l(e_0)^3 + l(e_0)^2 (d_1+d_2) + \frac{1}{2}l(e_0)\left(d_1^2 +d_2^2 \right). 
\end{align}
It is evident that $\Delta(e_{0};R)\geq \Delta(e_{0};R^{'})$. Hence proved.
\endproof

Based on Lemma \ref{lem:If-an-edge}, we further develop a lower bound of the AoI characterized by another virtual route, which eliminates the dependence on $d_1$ and $d_2$, as shown in the following lemma.

\begin{figure}[t]
\centering \includegraphics[scale=0.45]{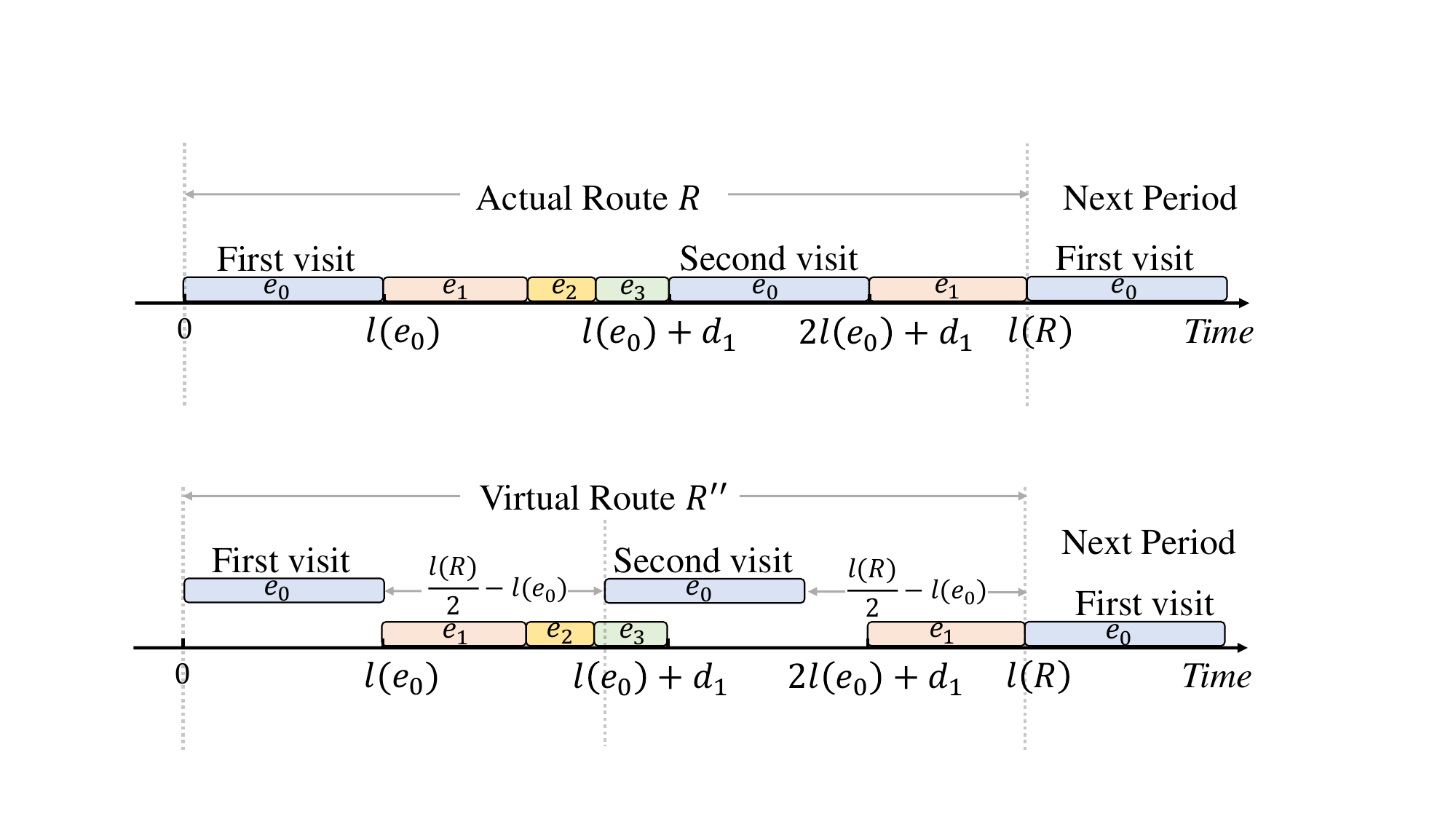}

\caption{An example of virtual route $R^{''}$. $R^{''}$ has the same visiting sequence of edges as the actual route $R$. The distance between two visits of $e_0$ is $\frac{l(R)}{2}-l(e_0)$. The virtual route $R^{''}$ is an infeasible route in this case as the second visit of $e_0$ overlaps with the traversal of $e_3$.  \label{fig:An-example-of}}
\end{figure}

\begin{lem}
\label{lem:If-an-edge-1}For a graph $\mathcal{G} = (V, E)$, if an edge $e_{0}\in E$ is traversed twice
within a route $R$, then the AoI of edge $e_0$ in $R$ is lower bounded by
that in the virtual route $R^{''}$, i.e., $\Delta(e_0;R^{''})\leq \Delta(e_0;R)$. In this virtual route $R^{''}$, the two visits of edge $e_{0}$
are from the same direction with the distance between two visits being $\frac{l(R)}{2}-l(e_0)$. Other edges under $R^{''}$
are traversed following the same sequence and direction as in $R$.
\end{lem}
\proof
By Lemma \ref{lem:If-an-edge}, we only need to prove that $\Delta(e_{0};R^{'})\geq \Delta(e_{0};R^{''})$.
We still consider the edge $e_{0}$ is visited twice within $R^{'}$
with distances $d_{1}$ and $d_{2}$ in between. We consider the following
optimization problem: 
\begin{eqnarray*}
\min_{d_1,d_2\geq0} & \Delta(e_{0};R^{'})\\
\mbox{s.t.} & d_{1}+d_{2} & =l(R)-2l(e_{0}).
\end{eqnarray*}
Substituting the constraint into the objective function, we have 
\begin{eqnarray*}
\Delta(e_{0};R^{'}) & = & l(e_{0})^{3}+l(e_{0})^{2}\Big(l(R)-2l(e_{0})\Big)\\
& + & \frac{1}{2}l(e_{0})\Big(d_{1}^{2}+(l(R)-2l(e_{0})-d_{1})^{2}\Big).
\end{eqnarray*}
Taking the derivative of the above equation over $d_1$, one can then easily obtain the optimal solution as $d_1=d_2=\frac{l(R)}{2}-l(e_{0}),$ which can be achieved by the route $R^{''}$ with $\Delta(e_{0};R^{''})  =  \frac{1}{4}l(e_{0})l(R)^2.$
Hence proved. 
\endproof
Note that the virtual route $R^{''}$ may not be a feasible route
in reality. Figure \ref{fig:An-example-of} provides an example of
the virtual route $R^{''}$, where the visits of edge $e_{0}$
are evenly distributed within the period, and the visits to other edges follow the same sequence as in the original route $R$. Based on the virtual route $R^{''}$, we derive the lower bound of the AoI for a given route $R$ in the following lemma.

\begin{lem}
\label{lem:(Lower-bound)-Given}(Lower bound)  Given a graph $\mathcal{G}=(V,E)$ and a patrol route
$R\in\mathcal{F}_{1}$. Suppose $E_{i}\in E$ is the set of edges that are traversed $i$ times within $R$. Then the AoI for
route $R$ is lower bounded by $\frac{1}{2}l(E_{1})^{2}+\frac{5}{4}l(E_{1})\cdot l(E_{2})+\frac{1}{2}l(E_{2})^{2}$, where $l(E_i)$ is the total length of the edges in the edge set $E_i$. 
\end{lem}
\proof
From Lemma \ref{lem:If-an-edge-1} and the fact that $l(E_1)+2l(E_2)=l(R)$,
we have 
 \begin{eqnarray*}
\overline{\Delta}(\mathcal{G}; R) & = &  \frac{1}{l(R)}\Big(\sum_{e\in E_1}\Delta(e;R) + \sum_{e\in E_2}\Delta(e;R)\Big)\\
 & \geq & l(R) \left( \frac{1}{2} l(E_1) + \frac{1}{4} l(E_2)\right)\\
 & = & \frac{1}{2}l(E_{1})^{2}+\frac{5}{4}l(E_{1})\cdot l(E_{2})+\frac{1}{2}l(E_{2})^{2}.
 \end{eqnarray*}

\begin{figure}
\centering\includegraphics[scale=0.7]{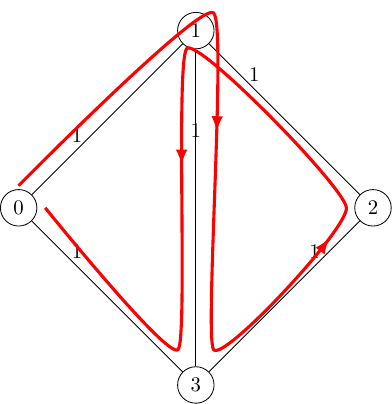}

\caption{Example of the lower bound in Lemma \ref{lem:(Lower-bound)-Given}.\label{fig:Example-of-the-lower}}
\end{figure}

This lower bound is tight in the example provided in Figure \ref{fig:Example-of-the-lower},
where the edges are of unit length and the route follows $(0,1,3,2,1,3,0)$. All the edges expect the edge $(1,3)$ are traversed once within the route.
Based on Algorithm \ref{alg:Compute-the-AoI}, the time-average AoI is computed as $\overline{\Delta}(R)=\frac{1}{2}l(E)^{2}+\frac{5}{4}l(E_{1})\cdot l(E_{2})+\frac{1}{2}l(E_{2})^{2}$, demonstrating that the bound is tight in this case.
\endproof

One can easily verify that Lemma \ref{lem:(Lower-bound)-Given} provides
a tighter lower bound than that in Lemma \ref{lem:For-any-connected}.
This lower bound will be useful for us to evaluate the performance
of approximation schemes later.

\subsection{Upper Bound of AoI in $\mathcal{F}_{1}$}\label{subsec:upper-bound}

Note that $\mathcal{F}_{1}$ contains various routes that traverse
all the edges at most twice. We now examine the largest AoI
resulting from routes within $\mathcal{F}_{1}$ by introducing the following
lemma. 
\begin{lem}
\label{lem:If-an-edge-2}For a graph $\mathcal{G} = (V, E)$, if an edge $e_{0}\in E$ is traversed twice
within a route $R$, then the AoI of edge $e_0$ in route $R$ is upper bounded by
that of the virtual route $\tilde{R}^{'}$, i.e., $\Delta(e_0;R)\leq \Delta(e_0;\tilde{R}^{'})$. In $\tilde{R}^{'}$, edge $e_{0}$
is traversed twice from the opposite directions, and other edges under 
are traversed following the same sequence and direction as $R$.
\end{lem}
\proof
It follows directly from Lemma \ref{lem:If-an-edge}. 
\endproof
Based on Lemma \ref{lem:If-an-edge-2}, we develop another bound in the following lemma. 
\begin{lem}
\label{lem:If-an-edge-3}For a graph $\mathcal{G} = (V, E)$, if an edge $e_{0}\in E$ is traversed twice
within a route $R$, then the AoI of edge $e_0$ in route $R$ is upper bounded by that
of the virtual route $\tilde{R}^{''}$, $\Delta(e_0;R)\leq \Delta(e_0;\tilde{R}^{''})$. In $\tilde{R}^{''}$, edges
are traversed following the same times and directions as $R$, the two visits of edge $e_{0}$
are adjacent and from opposite directions.
\end{lem}
\proof
We leave the proof to \ref{apen:A}.
\endproof

We next derive the upper bound of the time-average AoI for the entire graph based on Lemma \ref{lem:If-an-edge-3}.
\begin{lem}
\label{(Upper-bound)-Given}(Upper bound) Given a graph $\mathcal{G}=(V,E)$ and a route
$R\in\mathcal{F}_{1}$,
suppose $E_{i}\in E$ is the set of edges that are visited
$i$ times within $R$. Then the AoI for route $R$ is upper bounded
by $\frac{1}{2}l(E_{1})^{2}+\frac{3}{2}l(E_{1})\cdot l(E_{2})+l(E_{2})^{2}$. 
\end{lem}
\proof
We leave the proof to \ref{apen:B}.
\endproof

Lemma \ref{lem:If-an-edge-3} provides a useful bound for us to evaluate the approximation ratios of the algorithms that we will propose in the following subsections.

\subsection{The Duplicated Edge Scheme}\label{subsec:dup}

In this subsection, we introduce an approximation routing scheme based on a fundamental idea of duplicating the edges in the original graph, which we call the \emph{duplicated edge scheme}. We now provide the details of this scheme. As shown in Lemma \ref{lem:The-number-of}, for any graph $\mathcal{G}$,
one can always create a multigraph $\mathcal{G}^{'}$ by duplicating
the edges in $\mathcal{G}$. Eulerian cycles always exist in $\mathcal{G}^{'}$ since all the nodes in $\mathcal{G}^{'}$ have even degrees. Finding an Eulerian cycle within $\mathcal{G}^{'}$ takes a polynomial
time. One can simply apply Fleury's algorithm (with complexity $\mathcal{O}(n(E)^2)$) or Hierholzer's algorithm (with complexity $\mathcal{O}(n(E))$) to search for
an Eulerian cycle $R^{e}$ \cite{fleischner1990eulerian}.
If the UAV patrols the graph $\mathcal{G}$ along the route $R^{e}$, then
$R^{e}$ is a valid route within $\mathcal{F}_{1}$ as each edge in $\mathcal{G}$ is traversed twice. We can thus use this scheme as a polynomial-time approximation scheme for the AoI minimization problem,
which we summarize as Algorithm \ref{alg:Duplicated Edge Scheme}.

\begin{algorithm}
\footnotesize{
\begin{algorithmic}[1] \State{\textbf{Input:} the graph $\mathcal{G}=(V,E)$}
\State{Create a new multigraph $\mathcal{G}^{'}$ by duplicating
all the edges in $\mathcal{G}$} \State{Find an Eulerian cycle
$R^{e}$ in $\mathcal{G}^{'}$ by Fleury's or Hierholzer's algorithm}  
\State{\textbf{Output:} the route $R^{e}$} 
\end{algorithmic}
\caption{Duplicated edge scheme \label{alg:Duplicated Edge Scheme}}
}
\end{algorithm}

Since there are multiple Eulerian cycles within $\mathcal{G}^{'}$,
as shown in the proof of Lemma \ref{lem:The-number-of}, we denote all the periodic traversing
policies in Algorithm \ref{alg:Duplicated Edge Scheme} as $\mathcal{F}_{2}$. The
following theorem proves that following any Eulerian cycle $R\in \mathcal{F}_2$
as a traversing route can achieve a 2-approximation of the AoI.

\begin{thm}
\label{thm:dup} Denote all the periodic traversing
policies using Algorithm \ref{alg:Duplicated Edge Scheme} as $\mathcal{F}_{2}$, and $\mathcal{F}_{1}\subset\mathcal{F}$ be the set of policies defined in Definition \ref{def;F_1}. Then for all routes $R_2 \in \mathcal{F}_{2}$, it always holds that 
\begin{equation}\label{eq:ratio1}
\frac{\overline{\Delta}(\mathcal{G};R_{2})}{\overline{\Delta}(\mathcal{G};R_1^{*})}\leq\frac{l(E)^{2}}{\frac{1}{2}l(E)^{2}+\frac{1}{4}l(E_{1}^*)(l(E)-l(E_{1}^*))}\leq2
\end{equation}
and 
\begin{equation}
\frac{\overline{\Delta}(\mathcal{G};R_{2})}{\overline{\Delta}(\mathcal{G};R^{*})}\leq2,
\end{equation} 
where $R_1^*$ and $R^*$ denotes the optimal policies within the policy sets $\mathcal{F}_1$ and $\mathcal{F}$, respectively, and $E_1^*$ is the edges that are traversed once under $R_1^*$.
\end{thm}
\proof
Since all edges are duplicated in $\mathcal{G'}$, Eulerian cycle $R_2$ traverse all edges $E$ twice, that is $E_1 = \emptyset$ and $E_2=E$, we have $$\overline{\Delta}(\mathcal{G};R_{2}) \leq\frac{1}{2}l(E_{1})^{2}+\frac{3}{2}l(E_{1})\cdot l(E_{2})+l(E_{2})^{2}=l(E)^{2}$$ from Lemma
\ref{(Upper-bound)-Given}.
For any $R \in \mathcal{F}_1$, we have the lower bound from Lemma \ref{lem:(Lower-bound)-Given}:
\begin{eqnarray*}    
\overline{\Delta}(\mathcal{G};R^{*}_{1}) &\geq& \frac{1}{2}l(E^{*}_{1})^{2}+\frac{5}{4}l(E^{*}_{1})\cdot l(E^{*}_{2})+\frac{1}{2}l(E^{*}_{2})^{2} \\
&=& \frac{1}{2}l(E)^{2}+\frac{1}{4}l(E^{*}_{1})(l(E)-l(E^{*}_{1})).     
\end{eqnarray*}
Then the ratios are bounded by
\begin{eqnarray*}
\frac{\overline{\Delta}(\mathcal{G};R_{2})}{\overline{\Delta}(\mathcal{G};R^{*}_{1})} \leq \frac{l(E)^{2}}{\frac{1}{2}l(E)^{2}+\frac{1}{4}l(E_{1}^*)(l(E)-l(E_{1}^*))}\leq  2.  
\end{eqnarray*}
Similarly, using the conclusion in Lemma \ref{lem:For-any-connected}, we have 
$\frac{\overline{\Delta}(\mathcal{G};R_{2})}{\overline{\Delta}(\mathcal{G};R^{*})}\leq2.
$
We hence proved the theorem.
\endproof
Equation \eqref{eq:ratio1} shows that the approximation ratio between the duplicated edge scheme and the optimal route within $\mathcal{F}_1$ is always bounded by a factor determined by $E_1^*$. Although obtaining the explicit expression of $E_1^*$ can be challenging in many cases, we can draw several insights from this ratio. First, if $l(E_1^*)$ is significantly different from both 0 and $l(E)$ the ratio can be less than 2. Second, when $l(E_1^*)$ is close to $l(E)$, nearly all the edges are traversed once within the optimal route. In such scenarios, The duplicated edge scheme becomes inefficient since the majority of edges are traversed twice within each period. We will propose an algorithm in Section \ref{sec:CPP} that minimizes the distance of repeated traveling to resolve this issue.

Interestingly, although the duplicated edge scheme might seem inefficient, interestingly, traversing all the edges twice within a period can reduce the AoI in some cases. An example of this is provided in Appendix \ref{apen:C}.

\subsection{Approximation Scheme Based on the Chinese Postman Problem  (CPP) \label{sec:CPP}}

When an Eulerian cycle does not exist in the graph $\mathcal{G}$,
another intuitive approach is to create an Eulerian cycle by adding additional edges with the minimum distance, resulting in
a multigraph $\mathcal{G}^{''}$. This problem of finding a route that traverses all
the edges with the minimum route length is known as CPP. The CPP can be solved in polynomial time by the weighted
matching approach \cite{eiselt1995arc,edmonds1973matching}.

\begin{algorithm}
\footnotesize{
\begin{algorithmic}[1] \State{\textbf{Input:} the graph $\mathcal{G}=(V,E)$}
\State{Create a new  Eulerian multigraph $\mathcal{G}^{''} \supseteq \mathcal{G}$ by duplicating
the edges in $\mathcal{G}$ with the shortest total distance.}
\Statex{\Comment{This step can be handled using a weighted matching algorithm with complexity
$\mathcal{O}(n(V)^{3})$ \cite{edmonds1973matching}.}} \State{Find
an Eulerian cycle $R^{c}$ in $\mathcal{G}^{''}$ by Fleury's or Hierholzer's
algorithm} \State{\textbf{Output:} the route $R^{c}$.} \end{algorithmic} \caption{The CPP approximation scheme \label{alg:The-CPP-Approximation-Scheme}}
}
\end{algorithm}

Since all the nodes have even degrees within the multigraph $\mathcal{G}^{''}$,
the UAV can traverse $\mathcal{G}$ along an Eulerian cycle within $\mathcal{G}^{''}$.
We call this scheme the CPP approximation scheme, with details provided
in Algorithm \ref{alg:The-CPP-Approximation-Scheme}. Note that there
are also multiple Eulerian cycles within $\mathcal{G}^{''}$, so we
denote the periodic traveling policy along the Eulerian cycles within
$\mathcal{G}^{''}$ as policy set $\mathcal{F}_{3}$. In the following theorem, we provide the approximation ratio of the
policies within $\mathcal{F}_3$. 
\begin{thm}
\label{thm:cpp} Denote the policies by Algorithm \ref{alg:The-CPP-Approximation-Scheme} as $\mathcal{F}_3$, and $\mathcal{F}_{1}\subset\mathcal{F}$ be the set of policies defined in Definition \ref{def;F_1}. Then for all routes $R_3 \in \mathcal{F}_{3}$, it always holds that 
\begin{equation}\label{eq:ratio2}
\frac{\overline{\Delta}(\mathcal{G};R_3)}{\overline{\Delta}(\mathcal{G};R_1^{*})}\leq\frac{l(E)^{2}-\frac{1}{2}l(E)\cdot l(E_{1})}{\frac{1}{2}l(E)^{2}+\frac{1}{4}l(E_1^*)\cdot ( l(E)-l(E^*_{1}))}\leq2
\end{equation}
and 
\begin{equation}
\frac{\overline{\Delta}(\mathcal{G};R_{3})}{\overline{\Delta}(\mathcal{G};R^{*})}\leq\frac{l(E)^{2}-\frac{1}{2}l(E)\cdot l(E_{1})}{\frac{1}{2}l(E)^{2}}\leq2,
\end{equation}
where $R_1^*$ and $R^*$ denotes the optimal policies within the policy sets $\mathcal{F}_1$ and $\mathcal{F}$, respectively, $E_1^*$ is the edges that are traversed once under $R_1^*$, and $E_1$ is the edges that are traversed once under $R_3\in\mathcal{F}_3$.
\end{thm}
\proof
Since CPP minimizes the total route length within $\mathcal{F}_{1}$,
we know that the value of $l(R)=l(E_{1})+2l(E_{2})$ is minimized.
As $l(E)=l(E_{1})+l(E_{2})$, we know that CPP equals to the problem
of minimizing $l(E_{2})$. Let $E_{1}$ and $E_{2}$ be the edges
that traversed once and twice under CPP, and $E_{1}^{*}$ and $E_{2}^{*}$
be the edges that traversed once and twice under $R_{1}^{*}$, respectively,
we then have $l(E_{2})\leq l(E_{2}^{*})$ and $l(E_{1})\geq l(E_{1}^{*})$. 
From Lemma \ref{lem:(Lower-bound)-Given}, we have 
\begin{align*}
\overline{\Delta}(\mathcal{G};R_{1}^{*}) & \geq\frac{1}{2}l(E_{1}^{*})^{2}+\frac{5}{4}l(E_{1}^{*})\cdot l(E_{2}^{*})+\frac{1}{2}l(E_{2}^{*})^{2}\\
 & =\frac{1}{2}l(E)^{2}+\frac{1}{4}l(E_{1}^{*})(l(E)-l(E^{*}_{1})).
\end{align*}
We can then prove the results by Lemmas \ref{lem:For-any-connected}
and \ref{(Upper-bound)-Given}. 
\endproof

Although the explicit expression of $l(E_1^*)$ is generally intractable, we also gain the following insight from the ratio in Equation \eqref{eq:ratio2}. Note that the approximation ratio in Equation \eqref{eq:ratio2} can be far less than 2 when there are very few edges that need to be traversed twice under the CPP solution $\mathcal{G}^{''}$. For instance, if $l(E_2)\ll l(E_1)$, which implies $l(E_1)\approx l(E)$, the ratio $\frac{\overline{\Delta}(\mathcal{G};R_{3})}{\overline{\Delta}(\mathcal{G};R^{*}_{1})}\approx 1$ and $\frac{\overline{\Delta}(\mathcal{G};R_{3})}{\overline{\Delta}(\mathcal{G};R^{*})}\approx 1$. This demonstrates that the CPP approximation scheme can be more efficient than the duplicated edge scheme in scenarios when the number of edges requiring duplication in the CPP solution is minimal.

\begin{figure}[!t]
\centering \begin{subfigure}{.3\textwidth} \centering \includegraphics[width=1\linewidth]{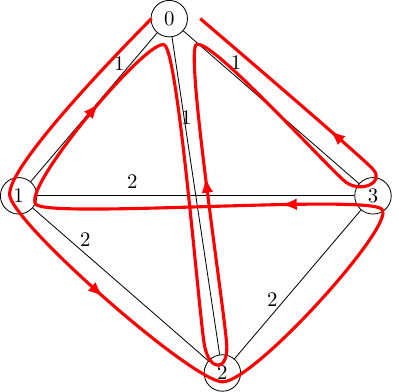}
\subcaption{Route 1} \end{subfigure} \begin{subfigure}{.3\textwidth}
\centering \includegraphics[width=1\linewidth]{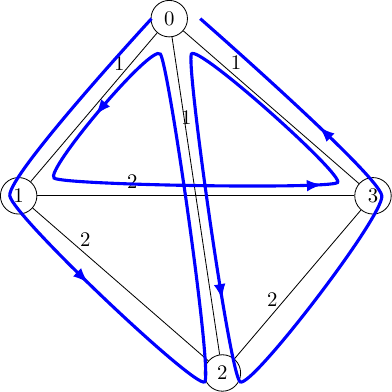}
\subcaption{Route 2} \end{subfigure} \caption{Example \ref{example:2}}
\label{Fig:Example2} 
\end{figure}

Although CPP can find the route with minimum distance to traverse all the edges, interestingly, the route for minimizing AoI is not necessarily a route that solves the CPP. That is, the optimal route in $\mathcal{F}_{1}$ may not
be a route in $\mathcal{F}_{3}$. An example is given in Appendix \ref{apen:D}. Moreover, since the policy set $\mathcal{F}_3$ contains multiple routes, the routes within $\mathcal{F}_{3}$ may result in distinct
AoI, as we illustrate in the following example.
\begin{example}
\label{example:2} In this example, we consider a graph with adjacency
matrix $A=\begin{pmatrix}0 & 1 & 1 & 1\\
1 & 0 & 2 & 2\\
1 & 2 & 0 & 2\\
1 & 2 & 2 & 0
\end{pmatrix}$. We consider two routes as follows: $R_1 = (0,1,2,3,1,0,2,0,3,0)$
and $R_2 = (0,1,2,0,1,3,0,2,3,0)$. A demonstrative graph of
these two routes is given in Figure \ref{Fig:Example2}. Both routes
have the same length of 12, while $R_1$ has an AoI of 49.333 and $R_2$
has an AoI of 45.778.   
\end{example}
Note that the two routes in Example \ref{example:2} result in distinct time-average AoI, with the AoI under $R_1$ being nearly $10\%$ higher than that under $R_2$. This difference illustrates that some Eulerian cycles can achieve significantly smaller AoI than others. The reason for this variance lies in the distribution of edge visits within each route. This observation motivates us to search for Eulerian cycles within $\mathcal{F}_2$ or $\mathcal{F}_3$ that achieve more balanced edge traversal. In the next section, we will explore methods to identify and select such Eulerian cycles, aiming to achieve better performance in terms of AoI.

\section{A Heuristic Approach and Numerical Studies\label{sec:heutistic-and-numerical}}
In this section, we develop
a heuristic approach that can select an Eulerian cycle with a relatively
smaller AoI compared to random selection. We provide a detailed discussion of the heuristic approach in Section \ref{subsec:heuristic}, and compare the approximation schemes numerically in Section \ref{subsec:numerical}.

\subsection{Heuristic Approach}\label{subsec:heuristic}

The general idea of the heuristic approach is as follows. First, we generate an undirected Eulerian multigraph $\mathcal{G}^{''}=(V,E^{''})$ by solving the CPP and duplicating the edges in $G$ with
the shortest distance. We then construct an Eulerian cycle on $\mathcal{G}^{''}$ using a modified Fleury's algorithm, where we aim to separate two visits of the same edge by assigning a potential to each visit. 

\begin{algorithm}[t]
\footnotesize{
\begin{algorithmic}[1] \State{\textbf{Input:} the graph $\mathcal{G}=(V,E)$}
\State{Create an undirected Eulerian multigraph
$\mathcal{G}^{''}=(V,E^{''})$ by duplicating the edges in $\mathcal{G}$ with the shortest total distance.}
\State{Initialize current node $v\leftarrow v_{0}$. The partial route
 $\bar{R}\leftarrow(v)$. Construct the graph $\bar{\mathcal{G}}=(V,\bar{E})\gets \mathcal{G}^{''}$ with $\bar{E} \gets E^{''}$} denoting edges not yet traversed
\While{$\bar{E}\neq\emptyset$} 
\State{$V_{\mbox{next}}\gets\emptyset$}
\For{$u\in V$} \If{$(v,u)\in\bar{E}$ \textbf{and} deleting $(v,u)$
not disconnect $\bar{E}$} \State{$V_{\mbox{next}}\gets V_{\mbox{next}}\cup u$}\Comment{Fleury's algorithm}
\EndIf \EndFor 
\For{$u\in V_{\mbox{next}}$} 
\If{$(v,u)$
appears once in $E^{''}$} 
\State{$p_{u}\gets\frac{1}{2}l(E^{''})$}
\ElsIf {$(v,u)$ needs two traversals \textbf{and} has been traversed in $\bar{R}$}
\State{$p_{u}\gets l((v,u)) + \tau$} 
\Else  \State{$p_{u}\gets\max\{\frac{1}{2}l(E^{''})+\epsilon,
l(\bar{R})+l((v,u))+\mbox{dist}(u,v_{0})\}$}
\EndIf \EndFor \State{ $u_{\mbox{max}}\gets\argmax_{u}p_{u}$}
\State{$\bar{E}\gets\bar{E}-(v,u_{\mbox{max}})$}
\State{Append $u_{\mbox{max}}$ to route $\bar{R}$}
\State{Moves to  $u_{\mbox{max}}$: $v\gets u_{\mbox{max}}$} 
\EndWhile
\State{\textbf{Output:} Route $\bar{R}$}
\end{algorithmic} \caption{The heuristic algorithm based on modified Fleury's algorithm  \label{alg:Heuristic}}
}
\end{algorithm}

We present the heuristic algorithm in Algorithm \ref{alg:Heuristic}, and explain the details in the following. After generating the multigraph $\mathcal{G}^{''}$ at Line 2, we apply the modified Fleury's algorithm to construct an Eulerian cycle (Line 3 to Step 25). In the modified Fleury's algorithm, we first construct a partial route $\bar{R}$ that starts from the source node $v_{0}$, and create a graph $\bar{\mathcal{G}}=(V,\bar{E})$ identical to $\mathcal{G}^{''}$, where $\bar{E}$ denotes the edges that have not yet been traversed. 
At each step, we select the next node to visit and add it to $\bar{R}$, until a complete route that traverses all the edges in $\mathcal{G}^{''}$ exactly once is accomplished. Suppose the partial route currently ends at node $v$, and $V_{\text{next}}$ is the set of eligible nodes to be visited from the current node $v$. Lines 5-10 of Algorithm \ref{alg:Heuristic} determine the set $V_{\text{next}}$. In a nutshell, $V_{\text{next}}$ contains all the nodes $u$ that are adjacent to the current node $v$ and where removing the corresponding eligible edge $(v,u)$ does not disconnect the graph $\bar{\mathcal{G}}$. After having the set $V_{\text{next}}$, our heuristic scheme selects the node $u$ from $V_{\text{next}}$ with the largest potential 
function $p_{u}$ (Line 11 to Line 20), and then removes this edge $(v,u)$ from the graph $\bar{\mathcal{G}}$ (Line 21). The partial route is updated by appending the selected node at the end (Lines 22 and 23). As this heuristic algorithm is based on Fleury's algorithm, its time complexity is $\mathcal{O}(n(E)^2)$.

The potential function, given in Equation \eqref{eq:4-1}, is the key part of the heuristic scheme, which we explain in the following.

\scriptsize{
\begin{eqnarray}\label{eq:4-1}
p_{u}  =  \begin{cases}
\frac{1}{2}l(E^{''}) & \mbox{if }(v,u)\mbox{ appears in } E^{''} \mbox{ only once but has not appeared in }\bar{R},\vspace{1mm}\\
l(e_{0})+\tau & \mbox{if } (v,u) \mbox{appears in } E^{''} \mbox{ twice and has been traversed in } \bar{R},\\
 & \mbox{where } \tau \mbox{ is the time elapsed since the last traversal of } (v,u); \vspace{1mm}\\
\max\{\frac{1}{2}l(E^{''})+\epsilon, & \mbox{if } (v,u) \mbox{ appears in } E^{''} \mbox{ twice but has not appeared in } \bar{R},\\
l(\bar{R})+l(e_{0})+\mbox{dist}(u,v_{0})\} & \mbox{where dist} (u,v_0) \mbox{ is the shortest distance from } u \mbox{ to } v_0, \\
&\mbox{ and } \epsilon \mbox{ is a small number}.
\end{cases}
\end{eqnarray}
}
\normalsize{}

By Lemma \ref{lem:If-an-edge} and letting $d_2=l(R)-2l(e_0)-d_1$, we know that if an edge $e_{0}$ is traversed twice from
the same direction within the route, its AoI can be expressed as:
\begin{align*}
\Delta(e_{0};R,d_{1}) & =  l(e_{0})^{3}+l(e_{0})^{2}\Big(l(R)-2l(e_{0})\Big)\\
& +\frac{1}{2}l(e_{0})\Big(d_{1}^{2}+(l(R)-2l(e_{0})-d_{1})^{2}\Big), 
\end{align*}
where $d_{1}$ is one of the distances between two visits of the edge.
Taking the derivative of AoI $\Delta(e_{0};R,d_{1})$ with respect to $d_{1},$ we have 
\begin{eqnarray*}
\frac{\Delta(e_{0};R,d_{1})}{\partial d_{1}}  =  l(e_{0})\big\{2d_{1}-l(R)+2l(e_{0})\big\}.
\end{eqnarray*}
The AoI of $e_{0}$ is thus decreasing in $d_{1}$ when $d_{1}\leq\frac{l(R)}{2}-l(e_{0})$
and increasing when $d_{1}>\frac{l(R)}{2}-l(e_{0})$. We observe the
same convexity for the AoI if edge $e_{0}$ is traversed twice from
opposite directions. Therefore, when an edge is traversed twice, we
want the time interval between the two visits to be as close to $\frac{l(R)}{2}-l(e_{0})$
as possible. If an eligible edge $e_{0}$ is supposed to be traversed
twice and has not been visited for more than $\frac{l(R)}{2}-l(e_{0})$
amount of time, we assign a high potential to it, since if it is not
traversed now, its average AoI will further increase. Conversely, If edge $e_{0}$
has just been visited and the time since the last visit is short, it is better to
select other edges first. If the edge $e_{0}$ is traversed only once in one period,
we can traverse it whenever it is eligible, as the sequence of traversing does not affect its average AoI. 

The potential function $p_{u}$
in Equation \eqref{eq:4-1} is thus designed based on this idea: We set the potential
of the edges that need to be traversed once as $\frac{1}{2}l(E^{''})$,
serving as a benchmark. For an edge $e_{0}$ that needs to be traversed
twice, if it has already been traversed once within the partial route $R^{'}$,
the time from the last visit is given as $\tau$, we set the potential as $l(e_0) + \tau$. This potential function value helps us prioritize edges that have not been visited recently. If $e_{0}$ has
not appeared in $\bar{R}$, we estimate the time from the last visit is at least 
$l(\bar{R})+l(e_{0})+\mbox{dist}(u,v_{0})$, which is the sum of the length
of the partial route $\bar{R}$, the length of edge $e_{0}$, and the
distance from $v_{0}$ to $u$. A large value indicates that edge $e_0$
should be visited soon to prevent a high AoI, while a smaller value suggests that it should still be prioritized over other single-visit edges. We thus set
the potential of edge $e_{0}$ as $\max\Big\{\frac{1}{2}l(E^{''})+\epsilon,l(\bar{R})+l(e_{0})+\mbox{dist}(u,v_{0})\Big\}$, where we let
$\epsilon=0.01$ be a small positive number to ensure that these edges are prioritized over the edges that need to be traversed only once.

\subsection{Numerical Study}\label{subsec:numerical}

\begin{figure*}[!t]
\centering 
\begin{subfigure}{.45\textwidth} \includegraphics[width=1\linewidth]{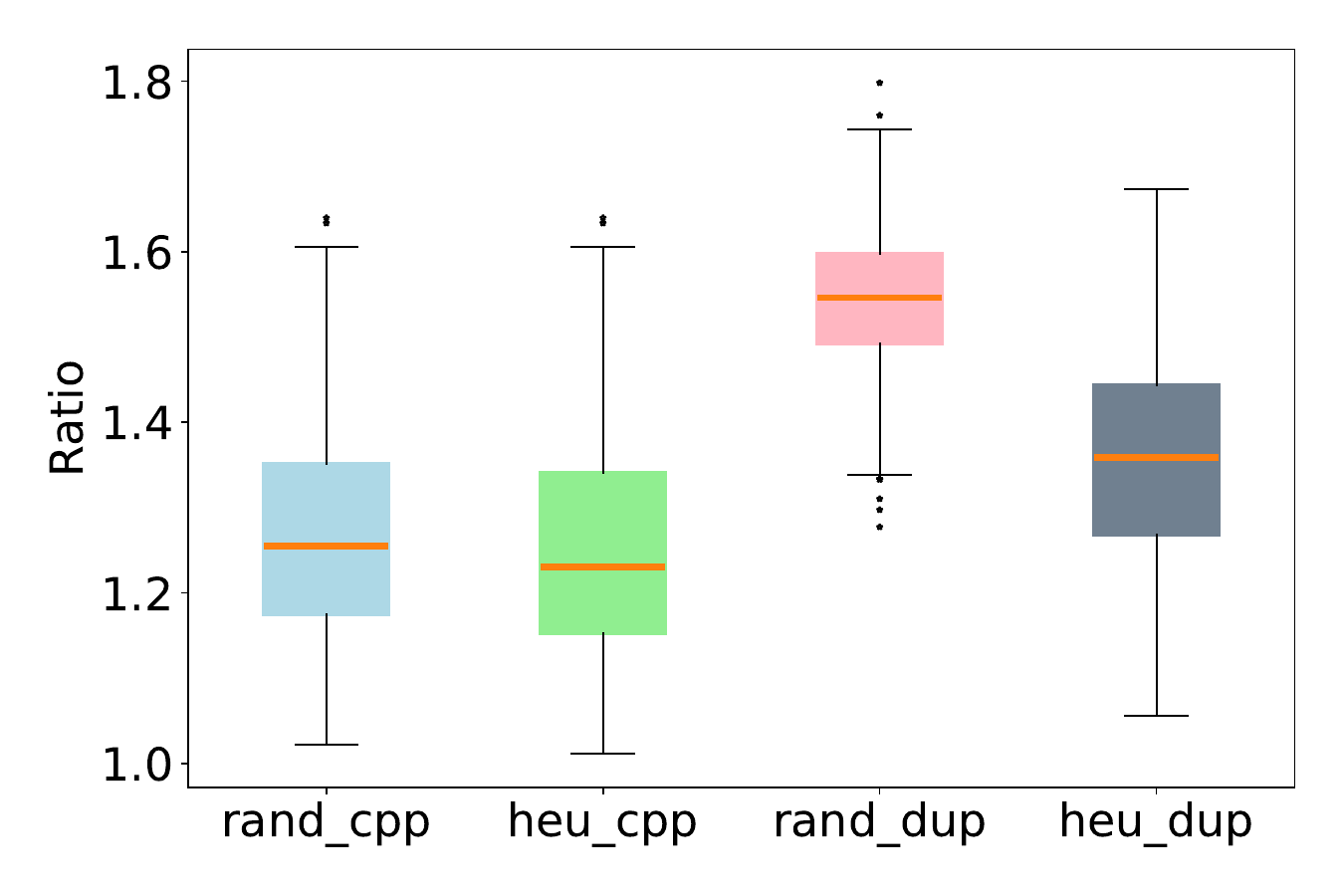}
\subcaption{$n=10,p=0.2$} \end{subfigure} 
\begin{subfigure}{.45\textwidth}
\includegraphics[width=1\linewidth]{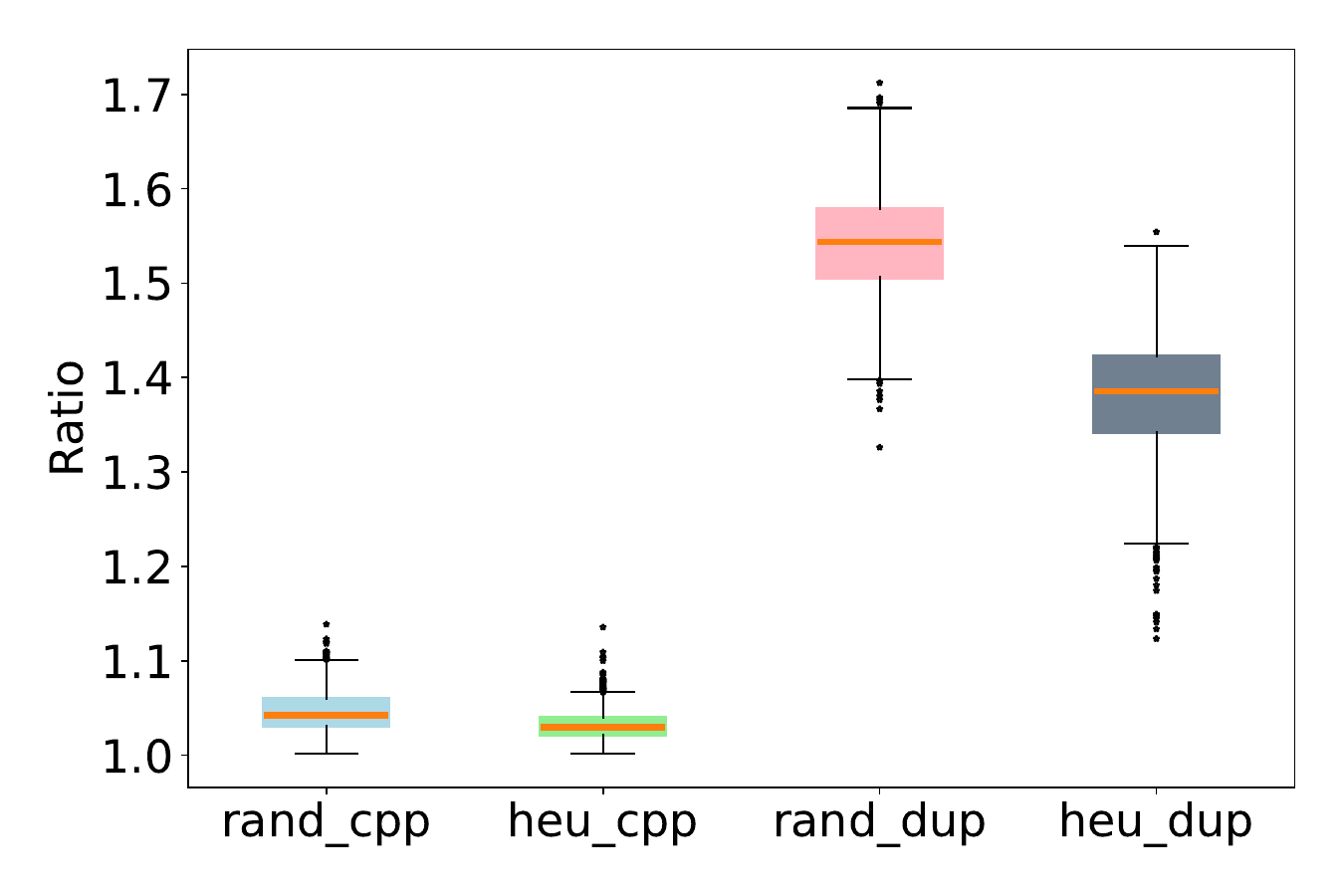}
\subcaption{$n=15,p=0.4$} \end{subfigure}
\begin{subfigure}{.45\textwidth}
    \includegraphics[width=1\linewidth]{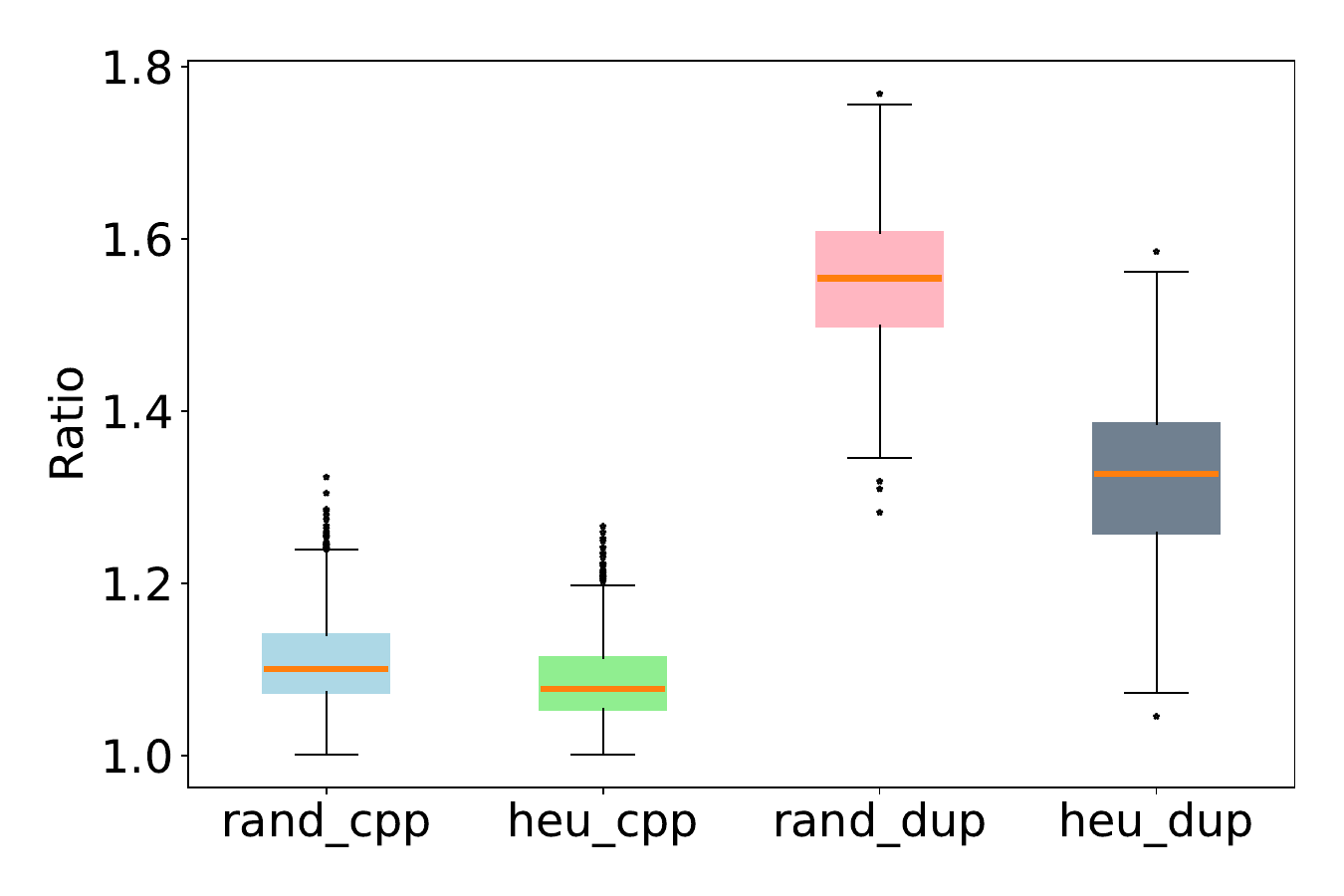}
\subcaption{$n=10,p=0.5$, planar graph} \end{subfigure}
\caption{Comparison of the algorithms}
\label{Fig:comparison} 
\end{figure*}

 We compare the average performance of algorithms through four case studies, as shown in Figure \ref{Fig:comparison}. Each subfigure in Figure \ref{Fig:comparison} is based on experiments conducted on graphs randomly generated by the Erdos-Renyi $G(n,p)$ model, with different number of nodes $n$ and probability $p$ of including an edge in the graph \cite{erdds1959random}. Specifically, we require the graph to be connected and non-Eulerian, and we retain 1,000 graphs that meet these criteria for each case study. In Figure \ref{Fig:comparison}(c), we also require the graph to be planar. The length of each edge over the graphs is randomly taken from the uniform distribution $U(0,10)$. 

The ``heu\_cpp'' denotes the heuristic algorithm provided in Algorithm \ref{alg:Heuristic}. The algorithm ``rand\_cpp'' constructs the route (Eulerian cycle) on the duplicated graph $\mathcal{G}^{''}$ obtained by solving CPP, using the Fleury's algorithm with each time selecting the next node from $V_{\text{next}}$ randomly. Note that we can also apply the heuristic selection of Equation \eqref{eq:4-1} on the duplicated graph $\mathcal{G}^{'}$ that is generated by simply duplicating all the edges within the original graph $\mathcal{G}$, denoted as ``heu\_dup''. Similarly, we apply random selection on the multigraph $\mathcal{G}^{'}$ and denote the algorithm as  ``rand\_dup''.
 
 The y-axis of each subfigure plots the ratio between the actual performance of each algorithm and the lower bound in Lemma \ref{lem:For-any-connected}. We observe that on both the multigraphs generated by duplication or CPP, the heuristic edge selection performs significantly better than random selection on average. Moreover, both the ``rand\_cpp'' and ``heu\_cpp'' algorithms exhibit better average performance than their counterparts on the duplicated graph $\mathcal{G}^{'}$. This indicates that searching for routes on the multigraph obtained by CPP can potentially achieve a better average performance for algorithms. 

\begin{figure*}[!t]
\centering 
\begin{subfigure}{.45\textwidth} \includegraphics[width=1\linewidth]{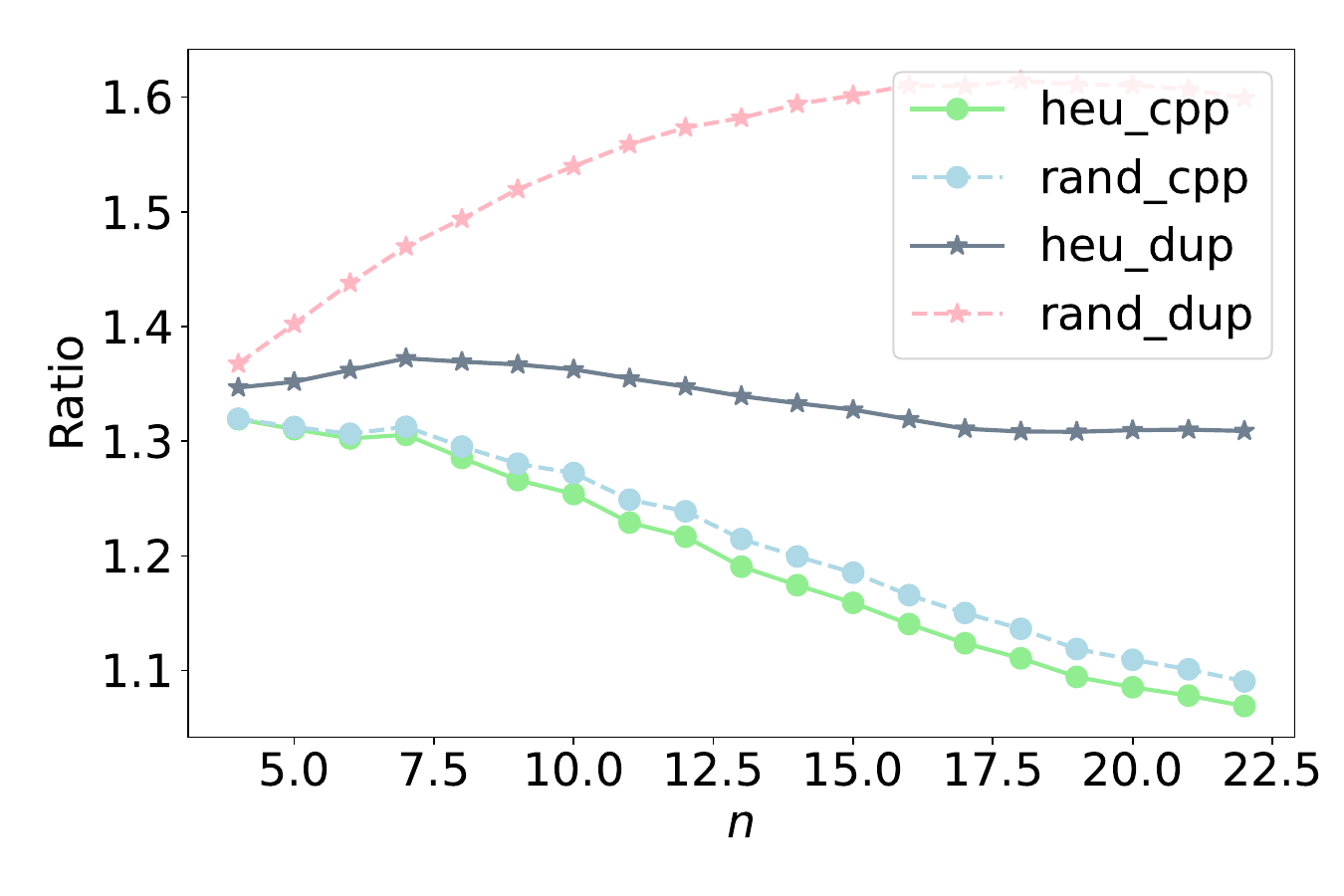}
\subcaption{$p=0.2$} \end{subfigure} 
\begin{subfigure}{.45\textwidth}
\includegraphics[width=1\linewidth]{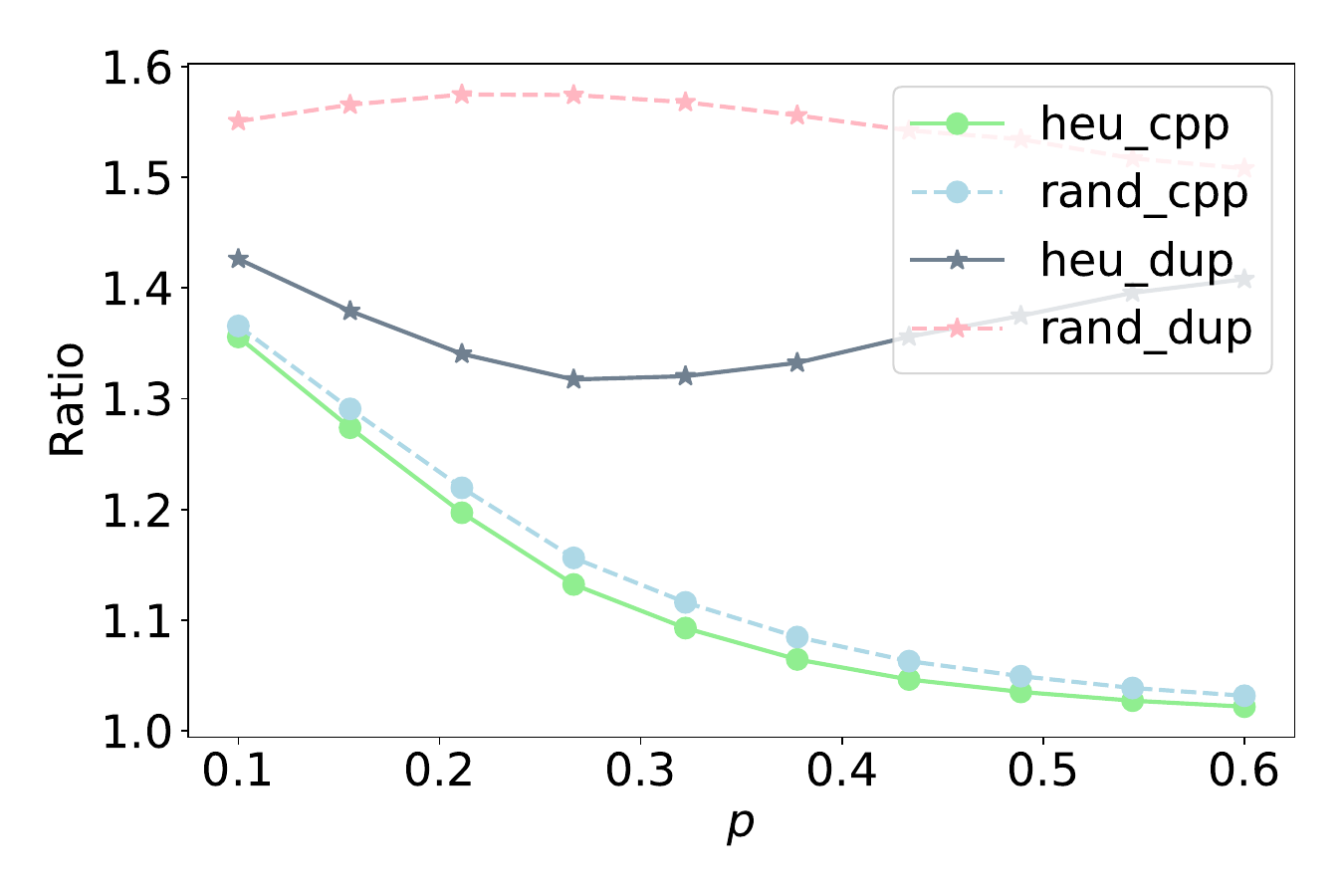}
\subcaption{$n=15$} \end{subfigure} 
\caption{Approximation ratios with different graph sizes and densities.}
\label{Fig:comparison2} 
\end{figure*}

We evaluate the robustness of the ``heu\_cpp'' algorithm in Figure \ref{Fig:comparison2}. In Figure \ref{Fig:comparison2}(a), we fix the probability parameter $p=0.2$ of the Erdos-Renyi model while varying the number of nodes $n$. In Figure \ref{Fig:comparison2}(b), we fix the number of nodes at $n=15$ and increase the probability $p$. An increase in $n$ results in a larger graph, while an increase in $p$ leads to a denser graph. Both graphs show that as either $n$ or $p$ increases, resulting in a graph with more edges, the advantages of using CPP-based policies become more pronounced.  This is because when the random graph contains more edges, the duplicated edge scheme becomes inefficient, as the optimal route might only need to traverse a small number of edges twice. Notably, the ``heu\_cpp'' algorithm consistently outperforms other algorithms across different graph sizes and densities, demonstrating its robustness and effectiveness. The heuristic selection ensures better distributions of edge visits, significantly reducing the average AoI compared to random selection strategies.

\section{Conclusions and Future Research\label{sec:conclusions-and-future}}
In this research, we investigate a routing problem
that aims to minimize the time-average AoI for
edges in general graphs. By formally defining the AoI of edges, we propose an algorithm to compute the AoI for any connected graph under a fixed route.   
We show that in the graphs without Eulerian cycles,
the optimal policy is difficult to determine, as the number of feasible routes is exponential. To address this, we provide the duplicated edge scheme and the CPP approximation scheme, both of which can achieve an approximation ratio of 2. Moreover, we introduce a heuristic scheme that maintains an approximation ratio of 2 but achieves a better average performance than randomly selecting the Eulerian cycles. Our numerical results demonstrate that the heuristic schemes significantly outperform random selection in various graph scenarios, and the CPP-based policies show increased efficiency in denser graphs.  

Our future research will focus on several key areas. We plan to explore more general routing policies beyond $\mathcal{F}_1$ set to improve flexibility and performance. Additionally, we aim to incorporate realistic constraints, such as battery limitations and charging times, into the AoI minimization problem to better reflect practical scenarios. Discussing the deployment of multiple UAVs in the graph patrol problem is another area of interest. Furthermore, we aim to provide a more rigorous proof for the complexity of the AoI minimization problem.

\section*{Acknowledgment}
This research is supported in part by National Natural Science Foundation of China under Grant
72301113, in part by Hubei Provincial Natural Science Foundation under grant 2023AFB857, and in part
by Fundamental Research Funds for the Central Universities, HUST: 2023WKYXQN001.

 \bibliographystyle{IEEEtran}
\bibliography{AoI_3}

{\appendix
\section{Proof of Lemma \ref{lem:If-an-edge-3}}\label{apen:A}
\proof
We focus on the virtual route $\tilde{R}^{'}$ and assume the distances between two visits of the edge $e_{0}$
is $d_{1}$ and $d_{2}$ respectively. Consider the following
optimization problem: 
\begin{eqnarray*}
\max_{d_1,d_2\geq0} & \Delta(e_{0};\tilde{R}^{'})\\
\mbox{s.t.} & d_{1}+d_{2} & =l(R)-2l(e_{0}).
\end{eqnarray*}
We substitute the constraint into the objective function, and rewrite the objective function $\Delta(e_{0};\tilde{R}^{'})$ by Equation \eqref{eq:lem3-1}
as 
\begin{align*}
\Delta(e_{0};\tilde{R}^{'}) &= {\frac{4}{3}}l(e_{0})^{3}+l(e_{0})^{2}\big(l(R)-2l(e_{0})\big)\\
&+\frac{1}{2}l(e_{0})\big(d_{1}^{2}+(l(R)-2l(e_{0})-d_{1})^{2}\big).
\end{align*}

One can easily prove that $\Delta(e_{0};\tilde{R}^{'})$ is a quadratic function of $d_1$ with axis of symmetry at $d_1=\frac{1}{2}(l(R)-2l(e_{0}))$. Given the constraints $0\leq d_{1}\leq l(R)-2l(e_{0})$, the 
the maximum value achieved at $d_1=0$ or $d_1=l(R)-2l(e_{0})$. In both cases, the two visits
of edge $e_{0}$ are adjacent. We thus have
\begin{align*}
\Delta(e_{0};\tilde{R}^{'}) \leq & \frac{4}{3}l(e_{0})^{3}+l(e_{0})^{2}\Big(l(R)-2l(e_{0})\Big) +\frac{1}{2}l(e_{0})\Big(l(R)-2l(e_{0})\Big)^2 \\
= & \frac{4}{3} l(e_0)^3 -l(R)l(e_0)^2 + \frac{1}{2} l(e_0) l(R)^2.
\end{align*}
Hence, we have proved the lemma. 
\endproof

\section{Proof of Lemma \ref{(Upper-bound)-Given}}\label{apen:B}
\proof
     From Algorithm \ref{alg:Compute-the-AoI} line 11 with the time interval between two adjacent visits $t=l(R)-l(e_j)$ for all $e_{j}\in E_{1}$. We thus have for $e_{j}\in E_{1}$ that 
\begin{align*}
\Delta(e_{j};R)  = &  \frac{1}{2}\left(l(R)-l(e)\right)^{2}\cdot l(e)+\left(l(R)-l(e)\right)\cdot l(e)^{2}+\frac{1}{2}l(e)^{3}  \\
= & \frac{1}{2} l(R)^2 l(e_j).
\end{align*}
From Lemma \ref{lem:If-an-edge-3}, we have for $e_{j}\in E_{2}$
that $\Delta(e_{j};R)\leq \frac{4}{3} l(e_j)^3 -l(e_j)^2 \cdot l(R) + \frac{1}{2} l(e_j) l(R)^2$.
Moreover, since in $R\in\mathcal{F}_{1}$, no edge is traversed for more than
twice, we have $l(E_{1})+2l(E_{2})=l(R).$ Thus, we can express the time-average AoI for the graph as: 
\begin{align}
\nonumber \overline{\Delta}(\mathcal{G};R)  & =  \frac{1}{l(R)}\Big(\sum_{e_{j}\in E_{1}}\Delta(e_{j};R)+\sum_{e_{j}\in E_{2}}\Delta(e_{j};R)\Big)\\ \nonumber
 & \leq  \frac{1}{2}l(R)\sum_{e_{j}\in E_{1}} l(e_{j})+ \frac{1}{l(R)}\Big(\frac{4}{3} \sum_{e_{j}\in E_{2}} l(e_j)^3 \nonumber\\
  &- l(R) \sum_{e_{j}\in E_{2}} l(e_j)^2 + \frac{1}{2} l(R)^2 \sum_{e_{j}\in E_{2}}l(e_j) \Big) \nonumber\\ 
 & =  \frac{1}{2}\Big[l(E_{1})+2l(E_{2})\Big]\cdot\Big[l(E_{1})+l(E_{2})\Big]\nonumber\\
  &+\sum_{e_{j}\in E_{2}}l(e_{j})^{2}\Big(\frac{4}{3}\frac{l(e_{j})}{l(E_{1})+2l(E_{2})}-1\Big). \label{eq:3-5}
\end{align}

To find an upper bound for the last term of the above equation, we assume set $E_2$ contains $m$ edges, denoted as $e_1, e_2, \dots, e_m$. Maximizing average AoI is equivalent to solving the following optimization problem: 
\begin{eqnarray*}
\max_{l(e_1),l(e_2),...,l(e_m)} & \sum_{j=1}^{m} l(e_j)^{2}\left(\frac{4}{3}\cdot \frac{l(e_j)}{l(E_{1})+2l(E_{2})}-1\right)\\
\mbox{s.t.} & \sum_{j=1}^{m} l(e_j)=l(E_{2}).
\end{eqnarray*}
The Lagrangian function of the optimization problem is:
\begin{align*}    
& L\Big(l(e_j), l(e_j), \dots, l(e_j), \lambda\Big) \\
 = & \sum_{j=1}^{m} l(e_j)^{2}\Big(\frac{4}{3} \cdot \frac{l(e_j)}{l(E_{1})+2l(E_{2})}-1\Big) - \lambda \Big( \sum_{j=1}^{m} l(e_j)-l(E_{2}) \Big).
\end{align*}
The KKT conditions $\partial L /\partial l(e_j)=0$ and $\partial L /\partial \lambda=0$ imply that: 
\begin{eqnarray*}
\frac{4l(e_j)^{2}}{l(E_{1})+2l(E_{2})}-2l(e_j)-\lambda & = & 0 \quad \forall j=1, \dots, m\\
\lambda\Big[\sum l(e_j)-l(E_{2})\Big]  &= & 0.
\end{eqnarray*}

So the optimal solution satisfies $l(e_1)=l(e_2)=\dots=l(e_m)$ and the optimal
value is given by: 
$$
g(m)=\frac{4 l(E_{2})^{3}}{3(l(E_{1})+2l(E_{2}))}\cdot \frac{1}{m^{2}}-l(E_{2})^{2}\cdot \frac{1}{m}.
$$

Taking the derivative of $g(m)$ with respect to $m$, we have: 
\begin{eqnarray*}
\frac{\partial g(m)}{\partial m} & = & \frac{4 l(E_{2})^{3}}{3(l(E_{1})+2l(E_{2}))}\cdot \Big(\frac{-2}{m^{3}}\Big)+\frac{l(E_{2})^{2}}{m^{2}}\\
 & = & \frac{l(E_{2})^{2}}{m^{2}}\left(1-\frac{8l(E_{2})}{3(l(E_{1})+2l(E_{2}))}\cdot \frac{1}{m}\right).
\end{eqnarray*}
The function $g(m)$ thus has one stationary point at
$$
m^{*}=\frac{3}{8} \cdot \frac{l(E_{1})+2l(E_{2})}{l(E_{2})}, 
$$ 
and $g(m)$ decreases when $m \leq m^{*}$ and increases when $m\geq m^{*}$. The maximal value
of $g(m)$ is achieved at $m\rightarrow\infty$ or $m=1$, with $\lim_{m\rightarrow\infty}g(m)=0$ and 
\begin{align*}
g(1) &= \frac{4 l(E_{2})^{3}}{3(l(E_{1})+2l(E_{2}))}-l(E_{2})^{2}\\
&= \frac{-2l(E_{2})^{3} - 3l(E_1)l(E_2)^2}{3(l(E_{1})+2l(E_{2}))} < 0.
\end{align*}
Hence, we always have $g(m)\leq0$.
We then have from Equation \eqref{eq:3-5} that 
\begin{align}
\overline{\Delta}(\mathcal{G};R)& \leq\frac{1}{2}\Big[l(E_{1})+2l(E_{2})\Big]\cdot\Big[l(E_{1})+l(E_{2})\Big]\nonumber \\
&=\frac{1}{2}l(E_{1})^{2}+\frac{3}{2}l(E_{1})\cdot l(E_{2})+l(E_{2})^{2}.\label{eq:3-6}
\end{align}

\begin{figure}
\centering\includegraphics[scale=0.7]{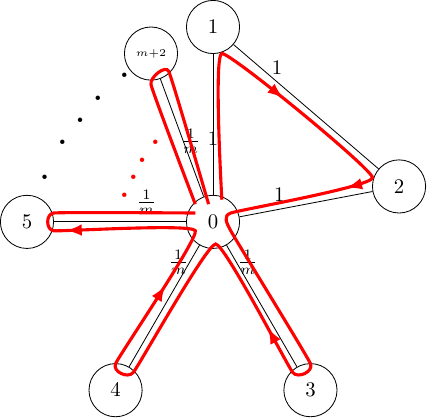}
\caption{Demonstrative example of the tightness of the upper bound\label{fig:Example-of-the}}
\end{figure}

The upper bound given in Equation \eqref{eq:3-6} is asymptotically tight in the example provided in
Figure \ref{fig:Example-of-the}. In the graph in
Figure \ref{fig:Example-of-the},  the edges $E_{0,1}$, $E_{1,2}$, and $E_{2,0}$ have the unit length, and the other edges all have the length $\frac{1}{m}$. The route takes $R_0 = (0,1,2,0,3,0,4,...,0,2+m,0)$, under which the edges $E_{0,1}$, $E_{1,2}$, and $E_{2,0}$ are traversed only once, while the other $m$ edges are traversed twice. We thus have $l(E_{1})=3$ and $l(E_{2})=1$. The length of the route
is $l(R_{0})=5$. The upper bound of the AoI is given as $10$ by Equation \eqref{eq:3-6}. For
edges $E_{0,1}$, $E_{1,2}$, and $E_{2,0}$, their AoI is identical. We can compute it by Algorithm \ref{alg:Compute-the-AoI} as
\begin{eqnarray*}
\Delta(E_{0,1};R_{0}) =  \Delta(E_{1,2};R_{0})=\Delta(E_{2,0};R_{0})= \frac{l(R_{0})^2}{2}.
\end{eqnarray*}
Using Lemma \ref{lem:If-an-edge-3}, the AoI for the other edges with $i\geq 3$ can be derived as:
$
\Delta(E_{i,0};R_{0}) =
  \frac{4}{3m^{3}}+\frac{l(R_{0})^{2}}{2m}-\frac{l(R_{0})}{2m^{2}}.
$
The time-average AoI for the entire graph is  given as: 
\begin{align*}
\overline{\Delta}(\mathcal{G};R_{0})= & \frac{1}{l(R_{0})} \Big(\Delta(E_{0,1};R_{0})+\Delta(E_{1,2};R_{0})+\Delta(E_{2,0};R_{0}) +\sum_{i=3}^{2+m}\Delta(E_{i,0};R_{0})\Big)\\
  = & \frac{3l(R_{0})}{2}+\frac{1}{l(R_{0})}\Big(\frac{4}{3m^{2}}+\frac{l(R_{0})^{2}}{2}-\frac{l(R_{0})}{2m}\Big)\\
  = & 2l(R_{0})+\frac{4}{3m^{2}l(R_{0})}-\frac{1}{m}.
\end{align*}

We then have $\lim_{m\rightarrow\infty}\overline{\Delta}(\mathcal{G};R_{0})=10$, which implies that the upper bound given by Equation \eqref{eq:3-6} is asymptotically tight as $m\rightarrow \infty$. 
\endproof

\section{Inefficiency of Traversing Fewer Edges}\label{apen:C}
\begin{figure}[t]
\centering \begin{subfigure}{.3\textwidth} \centering \includegraphics[width=1\linewidth]{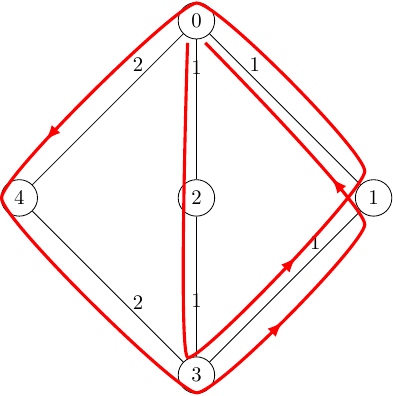}
\subcaption{$R_{1}$} \end{subfigure} \hspace{0.5cm} 
\begin{subfigure}{.3\textwidth}
\centering \includegraphics[width=1\linewidth]{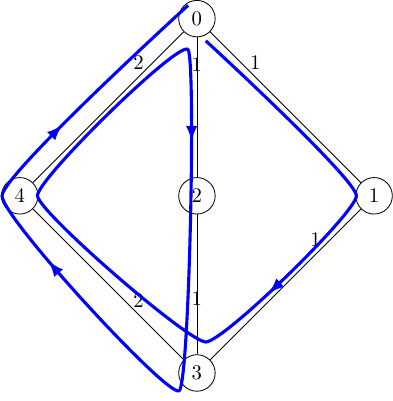}
\subcaption{$R_{2}$} \end{subfigure} \hspace{0.5cm} 
\begin{subfigure}{.3\textwidth}
\centering \includegraphics[width=1\linewidth]{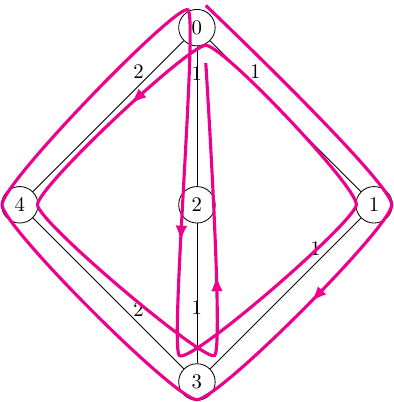}
\subcaption{$R_{3}$} \end{subfigure} \caption{Example \ref{example:1}}
\end{figure}
In the following example, we show that $\mathcal{F}_{2}$ can achieve
a smaller AoI than the routes that traverse some edges once and others
twice.
\begin{example}\label{example:1}
Consider a graph $\mathcal{G}$ with the adjacency matrix $A$ given as 
\[
A=\begin{pmatrix}0 & 1 & 1 & 0 & 2\\
1 & 0 & 0 & 1 & 0\\
1 & 0 & 0 & 1 & 0\\
0 & 1 & 1 & 0 & 2\\
2 & 0 & 0 & 2 & 0
\end{pmatrix},
\]
where each element $A_{i,j}$ in the matrix $A$ represents the edge length $l(E_{i,j})$. We consider
three routes as follows: $R_{1}=(0,2,3,1,0,4,3,1,0)$, $R_{2}=(0,1,3,4,0,2,3,4,0)$,
and $R_{3}=(0,1,3,4,0,2,3,1,0,4,3,2,0)$. Demonstrative graphs for the three routes are provided in Figure \ref{Fig:Example2}, where it is evident that
some edges within $R_{1}$ and $R_{2}$ are traversed only once, while
all the edges in $R_{3}$ are traversed twice. Using Algorithm \ref{alg:Compute-the-AoI}, we can obtain that $\overline{\Delta}(\mathcal{G};R_{1})=35.2$, $\overline{\Delta}(\mathcal{G};R_{2})=36$, and $\overline{\Delta}(\mathcal{G};R_{3})=33.67$. Thus, route $R_3$ achieves a smaller AoI than $R_1$ and $R_2$. The reason is that under Route $R_3$, the visits of edges are distributed relatively evenly. Although all the edges are traversed twice, the time periods between two visits of each edge are relatively short, keeping the AoI of each edge relatively small. In contrast, under $R_1$, edges $E_{0,4}$, $E_{0,2}$, $E_{2,3}$, and $E_{3,4}$ are visited once in a cycle, resulting in a large AoI for these edges. Similarly, edges $E_{0,1}$ and $E_{1,3}$ are rarely visited under $R_2$. This example demonstrates the importance of evenly distributing visits across all edges to minimize the time-average AoI for the entire graph. It suggests that routes ensuring frequent and evenly spaced visits to all edges can significantly reduce the overall AoI. This insight can be crucial when designing patrol routes for practical applications such as surveillance or maintenance tasks.

\end{example}

\section{Non-optimality of the CPP solution}\label{apen:D}
In the following example, we
show that the patrol route that minimizes AoI is not necessarily based
on a CPP solution.
\begin{example}
\label{example:3} In this example, we consider a graph with the adjacency
matrix
\[
A  =  \begin{pmatrix}0 & 1 & 1 & 1 & 1 & 1\\
1 & 0 & a & 0 & 0 & a\\
1 & a & 0 & a & 0 & 0\\
1 & 0 & a & 0 & a & 0\\
1 & 0 & 0 & a & 0 & a\\
1 & a & 0 & 0 & a & 0
\end{pmatrix}
\]
with $a=2.01$. We consider the following two routes: $R_1 = (0,1,2,0,3,4,0,5,1,0,2,3,0,4,5,0)$
with length 20.05, and  $R_2 = (0,1,2,0,3,4,0,5,1,2,3,4,5,0)$ with length
20.07. Figure \ref{Fig:Example3} provides a demonstrative
graph of the two routes. $R_1$ is the solution to CPP, with the route length $l(R_1) = 20.05$. Using Algorithm \ref{alg:Compute-the-AoI}, we find that $\overline{\Delta}(\mathcal{G};R_1)= 126.149$ and  $\overline{\Delta}(\mathcal{G};R_2)= 125.907$. Hence, in this case, the AoI minimizing route is not based
on the CPP solution. This example illustrates that the route achieving the minimum AoI does not always coincide with the route solving the CPP. While the CPP provides the shortest route for traversing all edges, the optimal route for minimizing AoI may require a different approach, highlighting the need to consider AoI-specific strategies.
\begin{figure}[!ht]
\centering \begin{subfigure}{.3\textwidth} \includegraphics[width=1\linewidth]{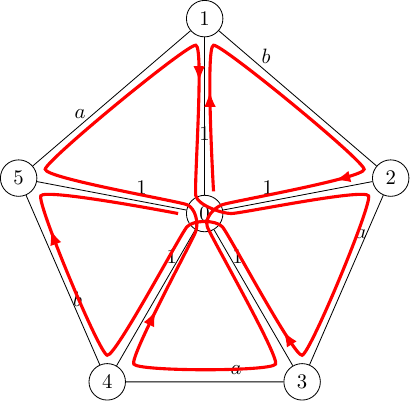}
\subcaption{Route 1} \end{subfigure} \begin{subfigure}{.3\textwidth}
\includegraphics[width=1\linewidth]{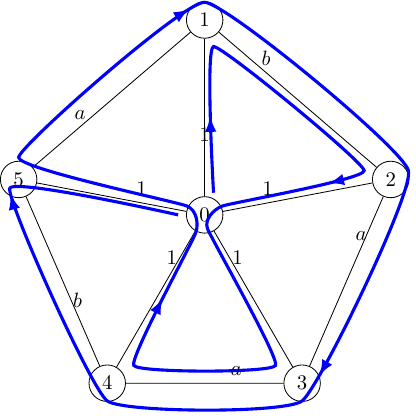}
\subcaption{Route 2} \end{subfigure} \caption{Example \ref{example:3}}
\label{Fig:Example3} 
\end{figure}
\end{example}

}

\end{document}